\documentclass[aps,twocolumn,graphics,floatfix,tightenlines]{revtex4-2}
\usepackage{graphics}
\usepackage{epstopdf}
\usepackage{epsfig}
\usepackage{graphicx}
\usepackage{epsf,epic}
\usepackage{color}
\usepackage{subfig}
\usepackage{amsmath}
\usepackage{booktabs}
\usepackage{multirow}
\usepackage{physics}
\usepackage{hyperref}
\usepackage{amsfonts}
\usepackage{wrapfig}
\usepackage{breqn}
\usepackage{pstricks}
\usepackage[utf8x]{inputenc}
\usepackage{multirow}
\usepackage{fancyref}
\usepackage{pst-node}
\usepackage{soul}
\usepackage{float}
\usepackage{bm}
\usepackage{dcolumn}
\newcommand{\etal}{\textit{et al.\ }}
\newcommand{\ie}{\textit{i.e.\ }}

\makeatletter
\usepackage{etoolbox} 
\appto{\appendix}{%
	\@ifstar{\def\theequation@prefix{A.}}%
	{}%
}
\preto\maketitle{%
  \begingroup\lccode`~=`,
  \lowercase{\endgroup
  \let\saved@breqn@active@comma~
  \let~}\active@comma 
}
\appto\maketitle{%
  \begingroup\lccode`~=`,
  \lowercase{\endgroup
  \let~}\saved@breqn@active@comma 
}
\makeatother
\begin{document}
\title{Quasiparticle band structure and excitonic optical response in V$_2$O$_5$
  bulk and monolayer}

\author{Claudio Garcia,$^1$ Santosh Kumar Radha,$^2$ Swagata Acharya,$^3$ and Walter R. L. Lambrecht$^1$}
\email{walter.lambrecht@case.edu}
\affiliation{$^1$ Department of Physics, Case Western Reserve University, 10900 Euclid Avenue, Cleveland, OH-44106-7079, USA}
\affiliation{$^2$ Agnostiq Inc. 325 Front St W, Toronto, ON M5V 2Y1, Canada}
\affiliation{$^3$ National Renewable Energy Laboratory, Golden, CO 80401, USA}

\begin{abstract}
  The electronic band structure of V$_2$O$_5$ is calculated using an all-electron 
  quasiparticle self-consistent (QS) $GW$ method, including electron-hole
  ladder diagrams in the screening of $W$, named  QS$G\hat W$
  and using a full-potential linearized muffin-tin-orbital basis set. 
  The optical dielectric function  calculated with the Bethe-Salpeter equation
  (BSE) exhibits excitons with large binding energy, consistent with spectroscopic
  ellipsometry data and other recent calculations using a pseudopotential plane wave based
  implementation of the many-body-perturbation theory approaches. Convergence
  issues are discussed. Sharp peaks in the direction perpendicular to the layers at high energy are found to be an artifact of the truncation of the
  numbers of bands included in the BSE calculation of the macroscopic dielectric function.  The static (electronic screening only) dielectric constant $\varepsilon_1(\omega=0)$  gives indices of refraction in good agreement with experiment. 
 The exciton wave functions are
  analyzed in various ways. They correspond to charge transfer excitons
  with the hole primarily on oxygen and electrons on vanadium, but depending on which exciton, the distribution over different oxygens changes. The dark exciton at 2.6 eV is the most localized and has the highest weight on the bridge oxygen, while the lowest bright excitons for in-plane polarizations at 3.1 eV for ${\bf E}\parallel {\bf a}$  and 3.2 eV for ${\bf E}\parallel{\bf b}$ have
  their higher weight on the chain and vanadyl oxygens. The exciton wave functions have a spread of about 5-15\AA, with asymmetric character for the electron distribution around the hole depending on which oxygen  the hole is fixed at.
    The same method applied first to bulk layered V$_2$O$_5$ is here
  applied to monolayer V$_2$O$_5$.
  The monolayer quasiparticle gap increases
  inversely proportional to interlayer distance once the initial  interlayer covalent couplings are removed which is thanks to the long-range nature of the self-energy and the
  reduced screening in a 2D system. The optical gap on the other hand is relatively independent of interlayer spacing because of the compensation between the
  self-energy gap shift and the exciton binding energy, both of which are proportional to the screened Coulomb interaction $\hat{W}$. Recent experimental
  results on very thin layer V$_2$O$_5$ obtained by chemical exfoliation
  provide experimental support  for an increase in gap. 
\end{abstract}
\maketitle

\section{Introduction}
Exciton binding energies in some layered transition metal oxides were recently found  to 
be extremely high, exceeding 1.0 eV.\cite{Radha21,Gorelov22} This is related to
the relatively low dispersion band edges in these materials and the low screening
of the Coulomb interaction in ionic materials, which suggest a Frenkel type exciton.
V$_2$O$_5$ is one such layered material
for which it was recently found that the excitons not only have strong binding energy
but for which these excitons nonetheless exhibit not so strongly localized
spatial extent and with an anisotropic  delocalization in unexpected directions.\cite{Gorelov22}

The band gap in V$_2$O$_5$ has  presented a puzzle for several years, since the first
$GW$ calculations were performed. While local density approximation (LDA) calculations\cite{Eyert98} gave results close to the experimentally accepted gap of about 2.3 eV,
which was extracted from Tauc plots of the optical
absorption,\cite{Kenny66} QS$GW$ calculations gave a much larger band gap
exceeding 4 eV.\cite{Bhandari15}
These results were also confirmed by other $G_0W_0$ implementations.\cite{Lany2013,vanSetten2017}
This puzzle was recently resolved by showing that
including electron-hole effects in the dielectric
function using the Bethe-Salpeter-Equation (BSE) approach \cite{Gorelov22} gives good agreement with
spectroscopic ellipsometry and reflectivity data.\cite{Mokerov1976,Losurdo2000}
These data show indeed sharp excitonic peaks with the lowest one at about 3.1 eV
for ${\bf E}\parallel {\bf a}$. The lower gap extracted from optical
absorption is still not completely understood and may either result from
excitons related to the indirect gap or phonon-mediated
activation of a dark exciton.\cite{Gorelov23}

While most $GW$ and BSE implementations are based on pseudopotential plane-wave
basis set implementations, all-electron implementations of many-body-perturbation theory
have recently become possible with linearized muffin-tin-orbital
and linearized augmented plane wave basis sets.\cite{Kotani07,questaalpaper,Friedrich10,Kutepov16,Kutepov17,Kutepov22,JiangHong16} The BSE approach was recently implemented using this approach by Cunningham \etal\cite{Cunningham18,Cunningham23}. An all-electron implementation is, in principle, preferable
since it avoids the uncertainties related to choosing pseudopotentials and describes
the core-valence exchange more accurately. Our first goal with the present paper
is to check whether similar strong excitons are obtained with an all-electron
BSE implementation and to further check the consistency  of the QS$GW$ band gap
between all-electron and pseudopotential based implementations.  Furthermore, 
in the usual QS$GW$ approach and also in  $G_0W_0$ approaches, $W$ is calculated in
the random phase approximation (RPA), meaning that the polarization propagator
is calculated as $P(1,2)=-iG(1,2)G(1,2)$ in terms of the Green's function
and is thus represented by a simple bubble diagram. (Here $1$ is short hand for $\{{\bf r}_1,\sigma_1,t_1\}$
including position, spin and time variable.) The screening is thereby underestimated
because it does not include electron-hole interaction effects. This has been recognized
for some time as a deficiency and has been corrected among other via an excursion into
time-dependent density functional theory, including a suitable exchange correlation
kernel in the calculation of the polarization propagator. Shishkin \etal\cite{Shishkin07}
used a kernel derived from BSE calculations, while Chen and Pasquarello\cite{ChenPasquarello15}
used the bootstrap kernel. Recently Cunningham \etal\cite{Cunningham23}
proposed an alternative method to include the ladder diagrams via a BSE formulation
in terms of the four-point polarization propagator. It can be viewed also as  a
vertex correction in the spirit of the Hedin equations.\cite{Hedin65,Hedin69}.
Unlike the approach of Kutepov \cite{Kutepov17,Kutepov22} who implemented similar vertex
corrections both in the screened Coulomb interaction $W=v+vPW$ with $P=-iG\Gamma G$ and the self-energy $\Sigma=-iGW\Gamma$,
and works directly toward implementing the Hedin equations self-consistently, the
approach of Cunningham uses the QS$GW$ approach, in which, in each iteration,
the full $G$ is replaced by $G_0$ corresponding to an updated Hermitian non-interacting Hamiltonian
$H_0$. The idea is to make the dynamic perturbation from $H_0$ as small as possible by incorporating
a static approximation of the self-energy into the exchange correlation potential of $H_0$.
The two Green's functions differ by
$G=ZG_0+\tilde G$ with $Z$ a quasiparticle renormalization factor and $\tilde G$
the incoherent part. But in $\Sigma$, $Z$ is then largely canceled by the vertex being
approximately proportional to $1/Z$, $\Gamma\propto1/Z$. 
This suggests that the vertex in $\Sigma$ should play a less important role in the QS$GW$
approach.\cite{Kotani07}  In practice, it gives accurate quasiparticle gaps and optical spectra when BSE is used for the latter without vertex corrections in the self-energy.\cite{Cunningham23,Radha21}
However, it has thus far been applied only to a limited number of materials.
It is thus of interest to test how it works for a challenging case like V$_2$O$_5$.

Finally, the question arises for such layered materials, whether the band gap
and optical properties will significantly change when going to the monolayer limit.
From Bhandari \etal\cite{Bhandari15} it is clear that in the LDA only a small
increase in gap occurs related to the breaking of some interlayer interactions
and hence reduced dispersion of the valence band edge.  However, in 2D materials,
one expects a strong reduction of the screening when the monolayer is isolated.\cite{Cudazzo11,Keldysh79} In Bhandari \etal\cite{Bhandari15} the QS$GW$ gap was shown to vary as $1/L$, with $L$ the interlayer distance, and
this led to an extremely large gap but which was of course overestimated.  Thus it becomes
of great interest to study how the inclusion of electron-hole interactions in the form
of ladder diagrams will affect the quasiparticle gap in a monolayer and how the
reduced screening will affect the exciton binding energies and exciton spectrum.
This is the second main goal of the present paper.

\section{Computational methods}
The density functional theory (DFT) calculations and subsequent  many-body-perturbation theory (MBPT)
calculations are performed using the {\sc Questaal} suite of codes as described in \cite{questaalpaper}.
These use a
full-potential linearized muffin-tin-orbital (FP-LMTO) basis set for the band structure calculations
and an auxiliary mixed interstitial-plane-wave and muffin-tin partial wave product basis set for the  representation of
two point quantities (bare Coulomb $v(1,2)$, screened Coulomb $W(1,2)$ and polarization propagator $P(1,2)$)
in the MBPT. The details of the Hedin-$GW$ implementation are given
in \cite{Kotani07} and for the Bethe-Salpeter-Equation approach in Cunningham \etal \cite{Cunningham18,Cunningham23}. Briefly, in the quasiparticle-self-consistent QS$GW$ method, a static and Hermitian
$\tilde\Sigma_{ij}=\frac{1}{2}\Re{\left[ \Sigma_{ij}(\epsilon_i)+\Sigma_{ij}(\epsilon_j)\right]}$
exchange-correlation potential 
is extracted from the energy-dependent $\Sigma_{ij}(\omega)$ where the matrices are given in the basis
$\psi_i$ of the $H_0$ Hamiltonian. The $\tilde\Sigma_{ij}-v_{xc}^{DFT}$  is then added to
the original $H_0^{DFT}$ and defines an updated $H_0$  from which the next $\Sigma(\omega)=-i G_0(\omega)\otimes W_0(\omega)$ is obtained, where the energy dependent self energy $\Sigma(\omega)$ is a convolution of the
one-particle Green's function and the screened Coulomb interaction. When iterated to self-consistency
in $\tilde\Sigma$, the quasiparticle energies become the same as the Kohn-Sham eigenvalues of the $H_0$
and the results are independent of the starting $H_0^{DFT}$. Here we use the generalized gradient
approximation (GGA) in the Perdew-Burke-Ernzerhof \cite{PBE} functional as starting DFT. 

The screened Coulomb interaction $W_0=(1-P_0^{RPA}v)^{-1}v$
is normally obtained from $P_0^{RPA}$ in the random phase approximation (RPA)
$P_0^{RPA}=-iG_0G_0$ (\ie using only the bubble diagram). The subscript $0$ indicates that it is calculated from the eigenstates of $H_0$. 
Instead, in  the QS$G\hat{W}$ method, the polarization propagator used is $P$, which
includes a summation over ladder diagrams instead of only the bubble diagram. 
This is done by converting to the four-particle generalized susceptibility $P$ and solving a
Bethe-Salpeter Equation and 
then converting back to the two-particle representation,
\begin{equation}
P(12)=P^{RPA}(12)-\int d(34) P^{RPA}(1134)W(34)P(3422).\label{eqPBSE}
\end{equation}
with $P^{RPA}(1234)=-iG(13)G(42)$. In practice this is done expanding the four-point quantities in
the basis set of single particle
eigenfunctions and amounts to diagonalizing an effective two-particle Hamiltonian.
It should be noted, however, that this involves solving a BSE at  a mesh of ${\bf q}$-points
because in $GW$ we need $W({\bf q},\omega)$. 
A static approximation is made for $W$  in Eq.(\ref{eqPBSE}) and the Tamm-Dancoff approximation (TDA) is made.  We note however, that this static, \ie $\omega=0$ approximation is only made in Eq.(\ref{eqPBSE})for $W$  but the frequency
dependence is maintained in the final $P$ through that of $P^{RPA}$.
Thus the final $W=(1-vP)^{-1}v$ used in the QS$GW$ equations is of course $\omega$ dependent. 

This approach is equivalent to adding a vertex correction to $P$ 
  in the Hedin set of equations as explained for example in \cite{Starke2012,Maggio2017}. 
  Details of the implementation and its justification are discussed in \cite{Cunningham23,Radha21}.

  Once the band structure is obtained in the QS$GW$ or QS$G\hat{W}$ approximations, from which we obtain the fundamental
  or quasiparticle gap, we can calculate the macroscopic dielectric function $\varepsilon_M(\omega)$
  for ${\bf q}\rightarrow0$.
  This involves another BSE equation with the kernel
  \begin{equation}
    K(1234)=\delta(12)\delta(34)\bar{v}-\delta(13)\delta(24)W,
  \end{equation}
    with $\bar{v}$ the microscopic part of the bare Coulomb interaction $v$, \ie  omitting the long-range
  ${\bf G}=0$ part in a Fourier expansion.  This is again done in the TDA and with a static $\hat{W}(\omega=0)$.  Here the first term in the kernel provides the local field corrections and the second provides the
    electron-hole interaction effects. Expanding this four-point quantity in the basis of one-particle eigenstates $\psi_{n{\bf k}}({\bf r})$,
    one obtains an effective two-particle Hamiltonian, given by
    \begin{eqnarray}
      H^{(2p)}_{n_1n_2{\bf k},n_1^\prime n_2^\prime{\bf k}^\prime}({\bf q})&=&\left(\epsilon_{n_2{\bf k}+{\bf q}}-\epsilon_{n_1{\bf k}}\right)\delta_{n_1n_1^\prime}\delta_{n_2n_2^\prime}\delta_{{\bf kk}^\prime}\nonumber\\
      &&-\left(f_{n_2{\bf k}+{\bf q}}-f_{n_1{\bf k}}\right) K_{n_1n_2{\bf k},n_1^\prime n_2^\prime{\bf k}^\prime}({\bf q})\nonumber \\
    \end{eqnarray}
    with $f_{n{\bf k}}$ the Fermi occupation function for band $n$ at ${\bf k}$.
    Diagonalizing this Hamiltonian in the Tamm-Dancoff approximation, where $n_1$ is restricted to be a valence state and $n_2$ a conduction band state,
    one obtains the exciton eigenvalues $E^\lambda({\bf q})$ and eigenvectors
    $A^\lambda_{n_1n_2{\bf k}}({\bf q})$.
    Introducing the shorthand $s=\{n_1n_2{\bf k}\}$, 
    the dielectric function is then given by
    \begin{eqnarray}
      \varepsilon_M(\omega)&=&1-\lim_{{\bf q}\rightarrow0}\frac{8\pi}{|{\bf q}|^2\Omega N_k}
        \sum_{ss^\prime}(f_{n_2^\prime{\bf k}^\prime+{\bf q}}-f_{n_1^\prime{\bf k}^\prime})\nonumber \\
        &&\rho_s({\bf q}) \sum_\lambda\frac{A^\lambda_s({\bf q})A^{\lambda*}_s({\bf q})}{E^\lambda({\bf q})-\omega\pm i\eta} \rho_{s^\prime}({\bf q})^*\label{epsmac}
    \end{eqnarray}
    with the matrix element
    \begin{equation}
      \rho_{n_1n_2{\bf k}}({\bf q})=\langle\psi_{n_2{\bf k}+{\bf q}}|e^{i{\bf q}\cdot{\bf r}}|\psi_{n_1{\bf k}}\rangle
    \end{equation}
Here, we have assumed no spin polarization and a factor two for spin and $N_k$ is the number of {\bf k}-points in the Brillouin zone. 
    The limit ${\bf q}\rightarrow0$ can be taken analytically, $e^{i{\bf q}\cdot{\bf r}}\approx 1+i{\bf q}\cdot{\bf r}$
    and then involves dipole matrix elements $\langle\psi_{n_2{\bf k}}| {\bf r}|\psi_{n_1{\bf k}}\rangle\cdot\hat{\bf q}$, where $\hat{\bf q}$ gives the
    direction along which we take the limit to zero and which corresponds to the polarization directions of the macroscopic tensor $\varepsilon_M(\omega)$.
    Finally, one converts the dipole matrix elements between Bloch states to velocity matrix elements divided by the band difference,
    $\langle \psi_{n_2{\bf k}}|[H,{\bf r}]|\psi_{n_1{\bf k}}\rangle=(\epsilon_{n_2{\bf k}}-\epsilon_{n_1{\bf k}})\langle\psi_{n_2{\bf k}}| {\bf r}|\psi_{n_1{\bf k}}\rangle$  and we then only
    need to diagonalize $H^{(2p)}({\bf q})$  at ${\bf q}=0$. 
    
    Besides shifts of the oscillator strength in the continuum, it
  can lead to bound excitons below the quasiparticle gap. The lowest bright excitons provide the
  exciton gap. At present, only direct dipole matrix elements are included between the one-particle states,
  so we only obtain
  direct excitons. Lower indirect excitons which would involve a phonon assisted transition could exist but
  are not calculated here. 
  The exciton eigenstates are a mixture of the vertical transition (between valence $v$ and
  conduction $c$ band states at a fixed {\bf k}), given by
  \begin{equation}
    \Psi^\lambda({\bf r}_h,{\bf r}_e)=\sum_{vc{\bf k}} A^\lambda_{vc{\bf k}} \psi_{v{\bf k}}({\bf r}_h)\psi_{c{\bf k}}({\bf r}_e)
  \end{equation}
  with ${\bf r}_h$, ${\bf r}_e$ the hole and electron position of the electron-hole pair bound in the
  exciton.  The summation over {\bf k} can lead to dark excitons if $A^\lambda_{vc{\bf k}}$ at 
  symmetry equivalent {\bf k} cancel each other even if the dipole matrix elements between these
  states are not zero at {\bf k}.
The coefficients $A^\lambda_{vc{\bf k}}$, which are the eigenvectors of the two-particle Hamiltonian   can be used to 
ascertain, which band-pairs $vc$ and at which {\bf k}-points contribute to a given exciton.
The exciton wavefunction modulo squared gives the probability to find the electron at position ${\bf r}_e$
for a fixed ${\bf r}_h$ or vice versa and is used to visualize the exciton spatial extent.

Further detail of the calculations are as follows. We use a $spdfspd$ basis set on V and O atoms,
meaning that two sets of Hankel function energy $\kappa^2$ and smoothing radius $R_{sm}$ are used
for the envelope functions of the LMTO basis set and with angular momenta up to $l=3$ for the first
set and $l=2$ for the second set. Inside the spheres, the basis functions are augmented by radial functions up to $l_{maxa}=4$ and V$_{3p}$ semi-core orbitals are treated as local orbitals, which means they are included in the basis set rather than in the core but have only an on-site contribution and are not augmented into other spheres.
We use slightly different muffin-tin spheres for the chemically different O-atoms optimized to avoid overlap between muffin-tin spheres. These are standard, well converged settings of the basis set.
The $GW$ self-energy matrix is calculated up to a maximum energy  of 2.56 Ry and approximated by an average
diagonal value above it as explained in \cite{Kotani07}.  Other details of the implementation, such as the construction of the mixed product basis set, which determines the dynamical screening, the contour integration approach for the self-energy and the offset $\Gamma$ method used to deal with the ${\bf q}\rightarrow0$
integrable divergence of $W$  all follow the approaches explained in \cite{Kotani07,Kotanijspj14}. 

\begin{figure}
  \includegraphics[width=8cm]{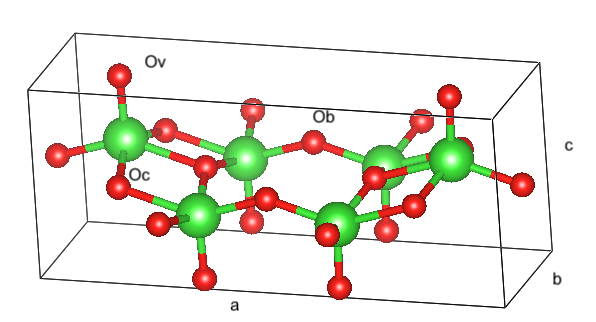}
  \includegraphics[width=5cm]{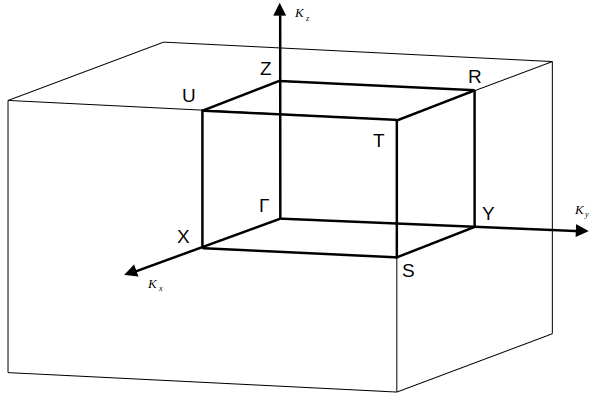}
  \caption{Crystal structure (top) and Brillouin zone (bottom)
    of V$_2$O$_5$. \label{figstruc}}
\end{figure}

The experimental structure in the $Pmnm$ space group is used as reported by Enjalbert and Galy\cite{Enjalbert86} and with lattice constants $a=11.512$ \AA, $b=3.564$ \AA, and $c=4.368$ \AA. The structure
is shown in Fig. \ref{figstruc} and consist of weakly van der Waals bonded layers, in the $c$
direction  and double zig-zag V-O chains in the $ab$-plane along $b$. The O in the chain are called chain  oxygen, O$_c$ and the chains are connected by bridge oxygens O$_b$.
The vanadyl oxygens O$_v$ are bonded to a single V and point alternating up and down along the chains
but in the same direction across a bridge  or rung of the ladders. This is called a ladder compound
with ladders consisting of the V-O$_b$-V rungs. Each V is surrounded by an approximately square pyramid
of 5 oxygens, one O$_v$, one O$_b$ and three O$_c$. The corresponding Brillouin zone labeling is also shown in Fig. \ref{figstruc}. 

\section{Results}
\subsection{Bulk}
\subsubsection{Band structure}
\begin{figure}
  \includegraphics[width=8cm]{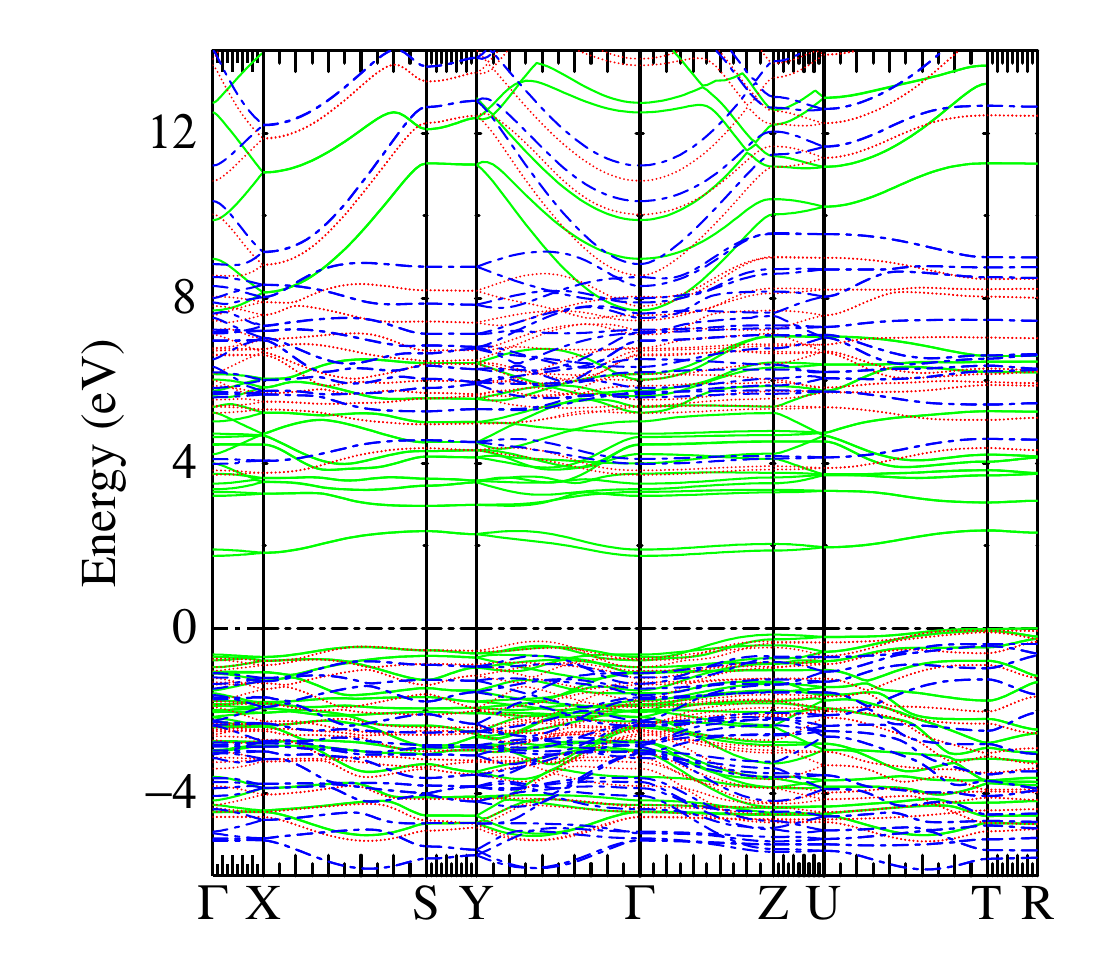}
  \caption{Band structure of V$_2$O$_5$ in GGA (green solid), QS$GW$ (blue dashed) and QS$G\hat{W}$ (red dotted)
    lines. The zero is placed at the VBM of GGA. \label{figbands}}
\end{figure}

First we show the band structure obtained in $GGA$, QS$GW$ and QS$G\hat{W}$ approximations in Fig. \ref{figbands}. The zero is placed at the valence band maximum (VBM) of the GGA. We can thus see how the $GW$ separately shifts valence bands
down and conduction bands up. This assumes the charge density is not changing too much between them.  We can see that the gap correction from GGA to QS$GW$ occurs primarily in the conduction band. When adding the ladder diagrams, the VBM
shifts back almost to the GGA position and the conduction band minimum (CBM) goes down
slightly. 
The gaps are given in Table \ref{tabgaps}. Our present QS$GW$ calculation differs from the older one by Bhandari\etal.\cite{Bhandari15,Bhandarithesis} in using a more complete basis set including $f$-orbitals. 
The CBM occurs at $\Gamma$, the VBM at T, the lowest direct gap at Z.
We give the indirect gap $\Gamma-T$, the lowest direct gap at $Z$ and the direct gap at $\Gamma$.
We can see that the inclusion of ladder diagrams reduces the gap correction beyond GGA by a factor $\sim0.77$,
close to a factor 0.8 as has been observed before for many other materials \cite{Bhandari18}.
Interestingly, our all-electron QS$G\hat{W}$ gaps agree closely with the QS$GW$ pseudopotential gaps
of Gorelov \etal\cite{Gorelov22}. This indicates that the screening of $W$ at the RPA levels is slightly underestimated in \cite{Gorelov22} compared to ours and this error is almost the same as the subsequent reduction of $W$ to $\hat W$ due to he inclusion of
electron-hole interactions via ladder diagrams. 

\begin{table}
  \caption{Band gaps in eV for bulk V$_2$O$_5$.\label{tabgaps}}
  \begin{ruledtabular}
    \begin{tabular}{lccc}
      method & indirect & minimum direct & direct at $\Gamma$ \\ \hline
      GGA    & 1.759    & 2.041 & 2.392 \\
      QS$GW$ & 4.370 & 4.799 & 5.075 \\
      QS$G\hat{W}$ & 3.781 & 4.178 & 4.452 \\
      QS$GW$-pseudo \footnote{From Gorelov \etal \cite{Gorelov22}} & 3.8 & & 4.4 \\
      $\frac{E_g(QSG\hat{W})-E_g(GGA)}{E_g(QSGW)-E_g(GGA)}$ & 0.774 & 0.775 & 0.768 \\
      QS$GW$\footnote{From Bhandari \etal.\cite{Bhandari15} and \cite{Bhandarithesis}}  & 4.00& 4.45 & 4.83 \\
    \end{tabular}
  \end{ruledtabular}
\end{table}

\begin{figure}
  \includegraphics[width=9cm]{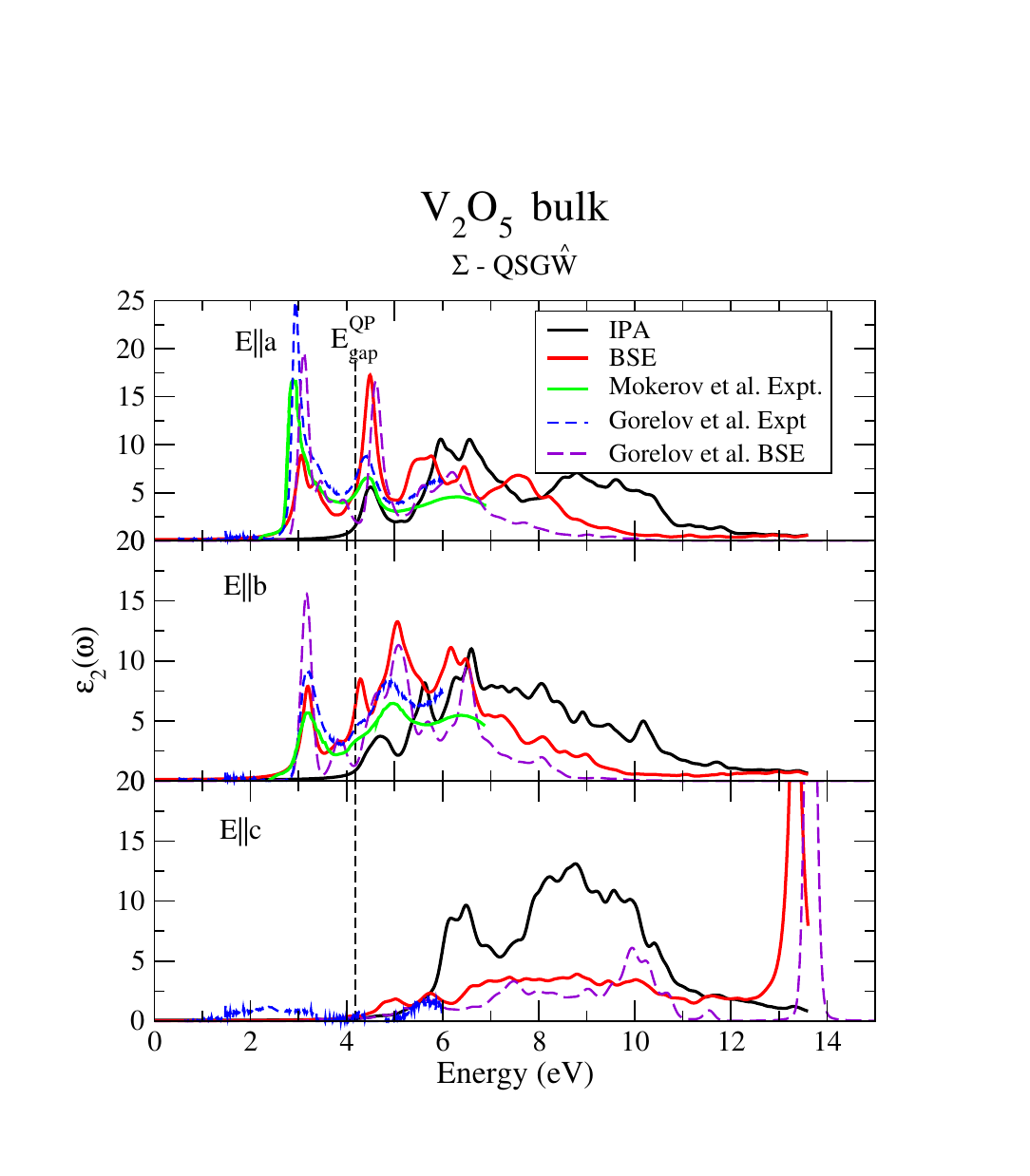}
  \caption{Imaginary part of the dielectric function for bulk V$_2$O$_5$ for three polarization,
    comparing IPA to BSE and to experimental results from Mokerov \etal\cite{Mokerov1976} and experimental and 
    calculated data from Gorelov \etal\cite{Gorelov22}. \label{figeps}}
\end{figure}

\subsubsection{Imaginary part of the dielectric function}
Next, in Fig. \ref{figeps} we show the macroscopic dielectric function  in the IPA and BSE both using
the QS$G\hat{W}$ bands. We can see that the BSE strongly alters the dielectric function and exciton peaks
with large binding energies appear significantly below the quasiparticle gap. This is in good agreement with
experimental results obtained from reflectivity by Mokerov \etal\cite{Mokerov1976} and more recent spectroscopic
ellipsometry results as shown in Gorelov \etal\cite{Gorelov22}.  We also show a comparison with
the BSE results by Gorelov \etal\cite{Gorelov22}.
The agreement is quite good, considering that a completely different code
is used in that work and that the intensities in the exciton region were found
to be quite sensitive to details of the calculation. 
Some of the differences with the work of Gorelov \etal\cite{Gorelov22} are  that our calculation
of the BSE two particle Hamiltonian includes 30 valence bands and 20 conduction bands, \ie all O-$2p$ and V-$3d$ related bands, while Gorelov \etal used 15 valence bands and 16 conduction bands. We used a $1\times5\times5$
k-point mesh (our $2\times6\times6$ k-mesh calculation is identical with $1\times5\times5$ k-mesh calculation with only $\sim$20 meV difference in band gap) in the BSE and $GW$ calculations while Gorelov \etal used
a $6\times6\times6$ grid. We used a smaller
number of {\bf k}-points in the $a$ direction because the unit cell is largest in this direction
and hence the Brillouin zone is smaller in this direction. Furthermore because of the anisotropy in structure, the
dispersion of the bands is larger in the $b$ than the $a$ or $c$ directions. Although it is difficult to fully trace the differences
between the pseudopotential plane wave calculation of Gorelov \etal \cite{Gorelov22} and the present calculation,
the main difference appears to lie in the calculation of the screened Coulomb interaction $W$ or the inverse dielectric function.
In the mixed product basis set representation $\varepsilon_{IJ}^{-1}({\bf q},\omega)$  the screening  on small spatial scales is
dealt with via products of partial waves in the spheres, which is easier to converge than by using high reciprocal  lattice vectors ${\bf G}$ in $\varepsilon^{-1}_{{\bf G}{\bf G}'}({\bf q},\omega)$.
The virtues of the product basis set were first introduced by Aryasetiawan and Gunnarsson \cite{Aryasetiawan94}. The idea  of describing the linear response using
partial wave fucntion inside the spheres  to avoid the need of high-lying band states was fruther elaborated in in recent work by Betzinger \etal \cite{Betzinger2012}.
Thus, we  guess that, in spite of the careful
convergence studies of the band gap as function of number of bands at the $G_0W_0$ level in \cite{Gorelov22}, their   RPA $W$ 
might still be slightly underestimated and agree better with our $\hat W$. However, given a certain value of
$W({\bf r},{\bf r}',\omega)$, which is in practice  represented by a different basis set, the two totally different codes
eventually agree on the quasiparticle band gaps resulting from it and the optical gap as obtained subsequently by the BSE.
This indicates, that the calculation of the $GW$ self-energy in terms of number of bands included as well as  the BSE calculaton 
are well converged in both approaches.
While \cite{Gorelov22} used a plasmon-pole approximation, it was tested against the more more rigorous
contour integration approach used in the {\sc Questaal} code and detailed in \cite{Kotani07}. While a pseudopotential treatment of the interaction with core and semi-core states is different from an all-electron approach,  it should be
noted that Gorelov \etal included V-$3s$ and V-$3p$ states as valence electrons. In our approach, V-$3s$ states are treated with atomic boundary conditions
at the muffin-tin sphere radii while V-$3p$ states are included as local orbitals. This means the $3p$ states are allowed to hybridize with the other valence states, but $3s$ states contribute only through their contribution to the total charge density, which affects the Hartree and exchange correlation potentials.
In contrast in a pseudopotential approach, the Hartree and exchange correlation potentials are obtained from the valence electron density only. However, the $3s$ states lie about 2 Rydberg below the $3p$ states which already lie at 3 Rydberg below the VBM. The core-treatment of the $3s$ electrons is thus certainly adequate. The effect of neglecting hybridisation of the $3p$ electrons was studied
in \cite{Bhandarithesis}. Treating V-$3p$ as core instead of local orbital
lowers the QS$GW$ gap by about 0.5 eV while it lowers the LDA gaps by only 0.2 eV. However, this is not an issue as we do include the V-$3p$ orbitals as local
orbitals and thus as valence electrons.  Thus we can safely conclude that these band structure atomic basis set aspects are adequately treated in both our and Gorelov \etal's calculation. The main difference thus lies in the basis sets used to represent the screening and the Hamiltonian.

For the purpose of visualizing the excitons, we subsequently used a 
$3\times5\times5$ mesh because this avoids overlapping exciton wave functions from the periodic images. However, this gives negligible differences in terms
of the energy spectrum itself.

In the polarization direction perpendicular to the layers ${\bf E}\parallel{\bf c}$  we may notice the strong suppression of BSE compared to IPA but also a
sharp peak at about 13.5 eV. This was not shown in Ref. \cite{Gorelov22}
because the energy scale was cut-off at lower energy but is also present in that calculation. It can also be seen in Ref. \onlinecite{Gorelov23} although somewhat less pronounced. The suppression of the imaginary part in the BSE compared to IPA in the energy range up to 10 eV or so, is a result of strong local field effects in layered systems~\cite{local-field}. This is the well-known classical depolarization effect. When a dielectric layer is placed in an external field, it induces a dipole  which produces a field opposite to the external field and this reduces the local field inside the layer by the dielectric constant~\cite{local-field1}. The sharp peak  at 13.5 eV, which lies just above the largest band to band transitions may also be related to local field effects.  As  discussed in
Cudazzo \etal \cite{Cudazzo2013} for 2D metals and also in an analysis of periodic boundary conditions artifacts in the modeling of local field effects by
Tancogne-Dejean \etal\cite{Tancogne-Dejean2015}, the imaginary part of
the dielectric function $\varepsilon_2(\omega)$ in systems with strong inhomogeneity can resemble the loss function $-{\rm Im}[\varepsilon^{-1}(\omega)]$.
This then explains both the suppression of the low energy region of the
dielectric function but also the occurrence of a plasmon like peak above the
energy range of the band pairs included. This feature is not present
in the independent particle approximation and this in itself indicates that it is a local
field effect.
  The  sharp peak we see here does not
quite look like a plasmon, because the latter is typically much broader. The
loss function was in fact calculated for V$_2$O$_5$ in Gorelov \etal\cite{Gorelov23}. However, as demonstrated in Supplemental Material \cite{supmat} the sharp peak
becomes suppressed when we include a higher number of conduction bands
$N_c=30$ instead of $N_c=20$. In fact, there are similar sharp features
at even higher energy and these become suppressed but a sharp feature then still occurs
at even higher energy. These in turn are reduced when adding  additional valence bands such as the deep lying O-$2s$-bands. This indicates that it is the
interaction of the sharp plasmon-like feature with the continuum of higher lying
electron-hole pairs that leads to a strong plasmon damping. The higher in
energy we want to obtain a converged $\varepsilon_2(\omega)$ the higher number
  of bands are needed in the BSE active space.

\subsubsection{Macroscopic dielectric constant from ${\bf q}\rightarrow 0$ limit}
In the results shown in Fig. \ref{figeps} the
limit ${\bf q}\rightarrow0$  of Eq. \ref{epsmac} is taken analytically, which requires to evaluate matrix elements of the commutator $[H,{\bf r}]$.
For a local potential, these amount to well-known momentum matrix elements.  However, for the QS$GW$ case, 
the evaluation of the commutator involves the non-local self-energy operator $\tilde\Sigma({\bf r},{\bf r}^\prime)$, which requires evaluating
$\nabla_{\bf k}\tilde\Sigma({\bf r},{\bf k})$ in which the ${\bf r}^\prime$ variable
is Fourier transformed to reciprocal space.\cite{DelSole93} Taking this derivative from the explicit expressions of the self-energy in terms of
the  LMTO  basis functions is cumbersome
and in the current implementation of the codes  
 involves some additional approximations, which
 experience has shown to lead typically to an overestimate of
 the matrix elements.
 Alternatively, we may consider directly the dielectric function at small but finite ${\bf q}$, which is obtained
 as part of the $GW$ procedure, 
 and extrapolate numerically to  ${\bf q}\rightarrow0$ along the three directions,
$\hat{\bf x}\parallel {\bf a}$, $\hat{\bf y}\parallel{\bf b}$ and $\hat{\bf z}\parallel{\bf c}$.
  This can then be done both at the
 RPA or the BSE level. This 
 allows us to more accurately evaluate $\varepsilon_1(\omega=0)$.
 This amounts to the static value but including only electronic, not phonon
 contributions, to screening, which is conventionally called $\varepsilon_\infty$. Experimentally, this corresponds to the index of refraction squared at a
 frequency well below the bands but also well above the phonon frequencies,
 which we can compare to experimental data by Kenny and Kannewurf,\cite{Kenny66}
 who obtained it by extrapolating the behavior of the index of refraction
 $n(\omega)$ for $\omega\rightarrow0$ in the region above the phonon bands.
 This provides an important test of the methodology because good agreement
 indicates that the QS$G\hat{W}$ method adequately describes dielectric screening. 

\begin{table}
  \caption{Indices of refraction $n=\sqrt{\varepsilon_1(\omega=0)}$ for different directions and in different approximations.\label{tabindex}}
  
  \begin{ruledtabular}
    \begin{tabular}{lccc}
      & $n_x$ & $n_y$ & $n_z$ \\ \hline
      RPA  & 1.88 & 1.83 & 1.75 \\
      BSE ${\bf q}\rightarrow0$ &  2.37 & 2.23 & 1.92 \\
      BSE ${\bf q}=0$ & 2.44 & 2.42 & 1.99 \\
      Expt.\footnote{Kenny and Kannewurf \cite{Kenny66}} & 2.07 & 2.12 & 1.97 \\
    \end{tabular}
  \end{ruledtabular}
  \end{table}

 Using finite small {\bf q} comes with its own set of numerical
 difficulties. It turns out  that to avoid unphysical behavior  such as
 negative values of $\varepsilon_2(\omega)$ it is necessary to replace the
 bare Coulomb interaction, $4\pi/q^2$, by a Thomas-Fermi screened
 $4\pi/(q^2+q_{TF}^2)$ with a small $q_{TF}$. We thus need to extrapolate both ${\bf q}\rightarrow0$ and $q_{TF}\rightarrow0$.
 Details of this procedure are given in Supplemental Material\cite{supmat}. The main results for the index of refraction are given in Table \ref{tabindex}.

We can see that the RPA calculated dielectric constants are systematically lower than the BSE calculated ones and the RPA is clearly seen to underestimate the experimental values.
 The BSE results obtained with the numerical extrapolation  are slightly
 smaller than the ones obtained with the analytically calculated matrix elements,
 which we call ${\bf q}=0$ instead of ${\bf q}\rightarrow0$.
 The values for $a$ and $b$ direction are close but  in inverse order
 from the experiment and larger than for the $c$ direction  but for the $c$
 direction the analytically calculated value is closer to the experiment than
 the numerical extrapolation while the opposite is true for the
 in-plane directions. This illustrates the  numerical difficulties with
 either approach. Nonetheless, overall the error in the indices of refraction is at most 15 \%.

 \subsubsection{Exciton analysis.}

 \begin{figure}
  \includegraphics[width=8.5cm]{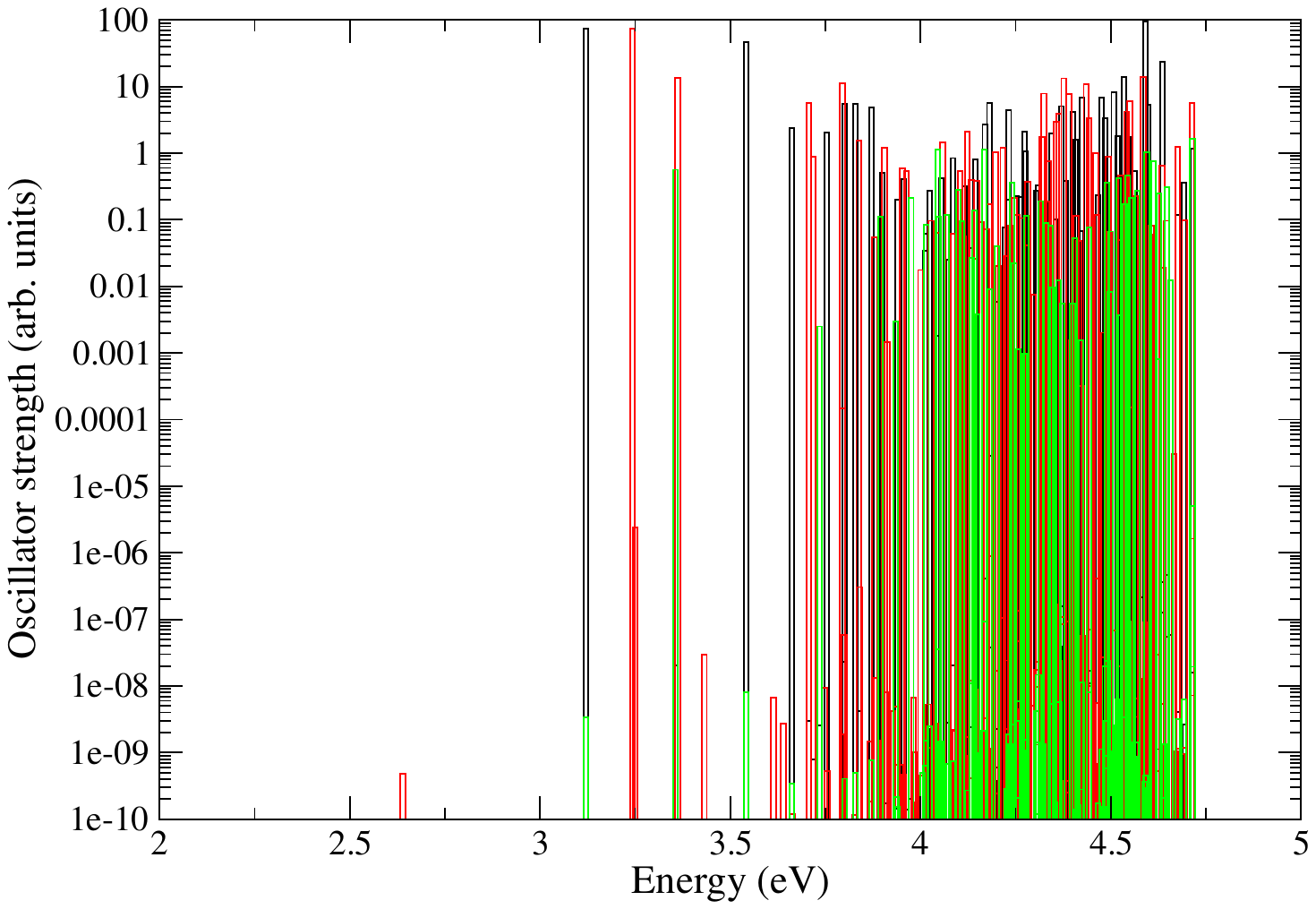}
  \caption{Eigenvalues of the two particle Hamiltonian with
    relative oscillator strengths on a log-scale. The colors indicate polarization:
    black, ${\bf E}\parallel{\bf a}$ , red ${\bf E}\parallel{\bf  b}$  and green ${\bf E}\parallel{\bf c}$.
   \label{fig2plevels}}
 \end{figure}
 
Besides the bright excitons, there are also several dark excitons.
An overview of the eigenvalues of the two-particle Hamiltonian
up to about the quasiparticle gap is shown in Fig. \ref{fig2plevels}
as a set of bar-graphs  with the oscillator strengths on a log-scale.
(Note that these were obtained with $N_v=30$ and $N_c=30$ but for these low energy excitons the results are equivalent for $N_c=20$.)
Any level  which has an oscillator strength lower than
0.1 may be considered dark as it is 1000 times smaller than the bright exciton oscillator strengths.
These oscillator strengths are not normalized and thus given in arbitrary units. Only their relative value is important here.
It is notable that an approximately doubly degenerate very dark exciton occurs well below the first bright excitons and 
near 2.6 eV. As was already mentioned in \cite{Gorelov22} these result from a destructive interference of the exciton eigenstates
at  symmetry equivalent {\bf k}-points rather than from zero dipole matrix elements at each individual {\bf k}.

\begin{figure*}
  (a) \includegraphics[width=5cm]{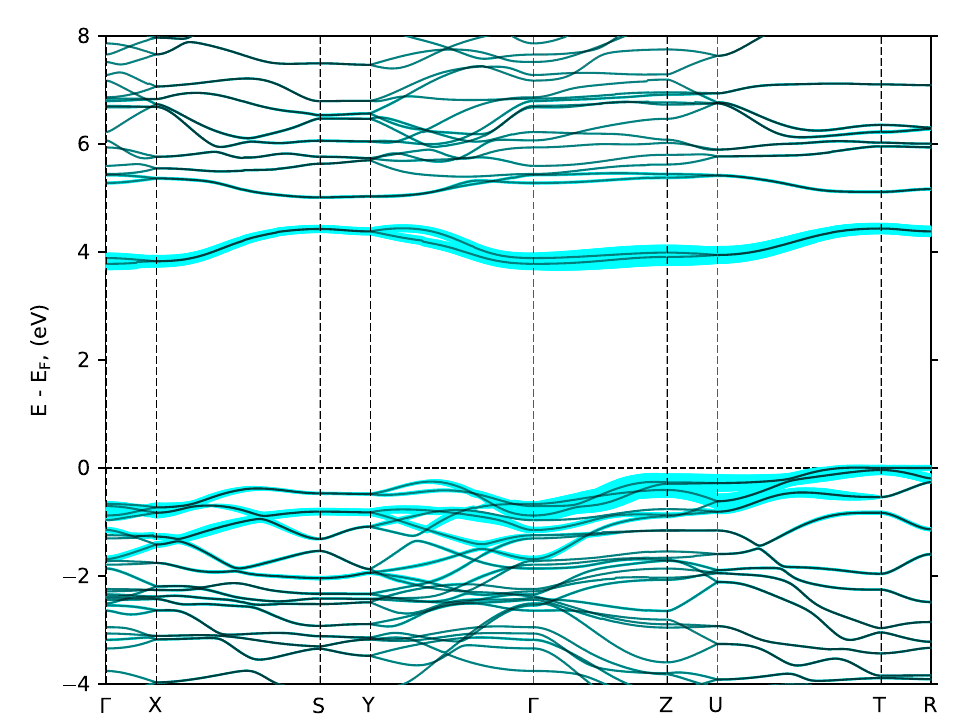}
  (b)  \includegraphics[width=5cm]{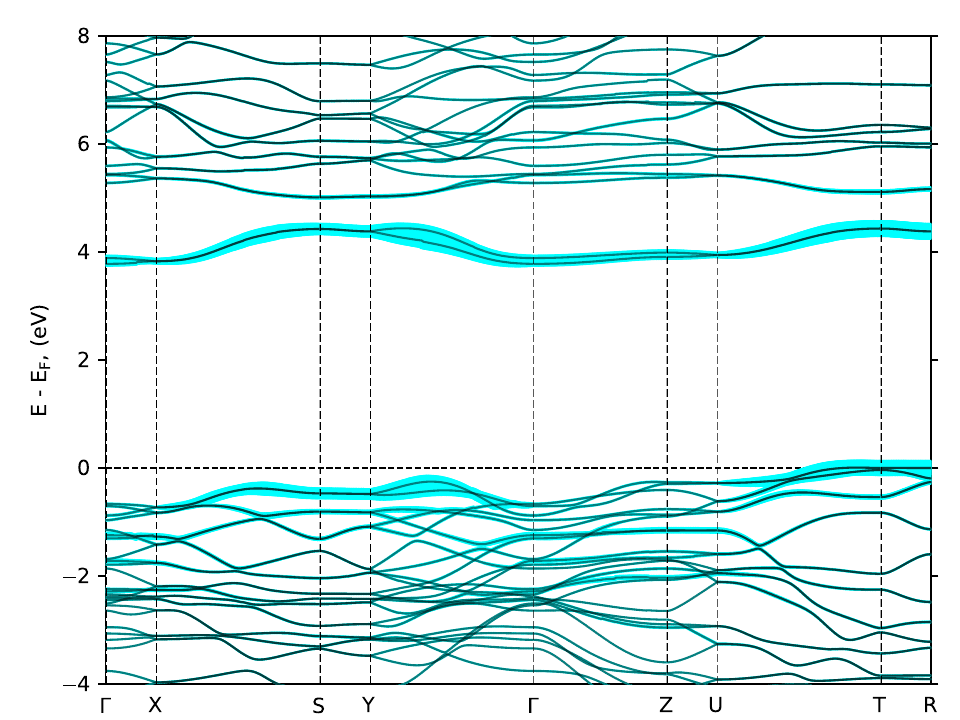}
  (c) \includegraphics[width=5cm]{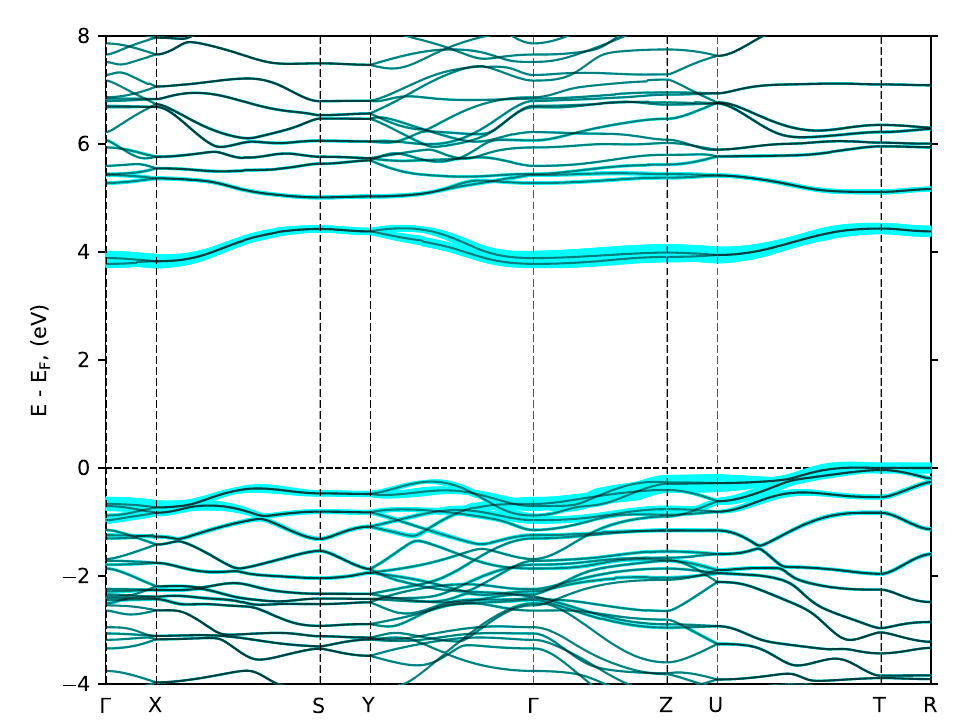} \\
  (d) \includegraphics[width=5.6cm]{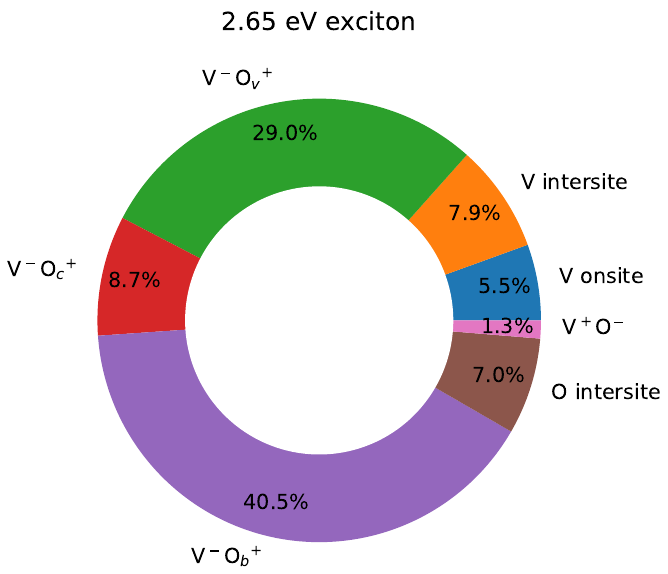}
  (e) \includegraphics[width=5.6cm]{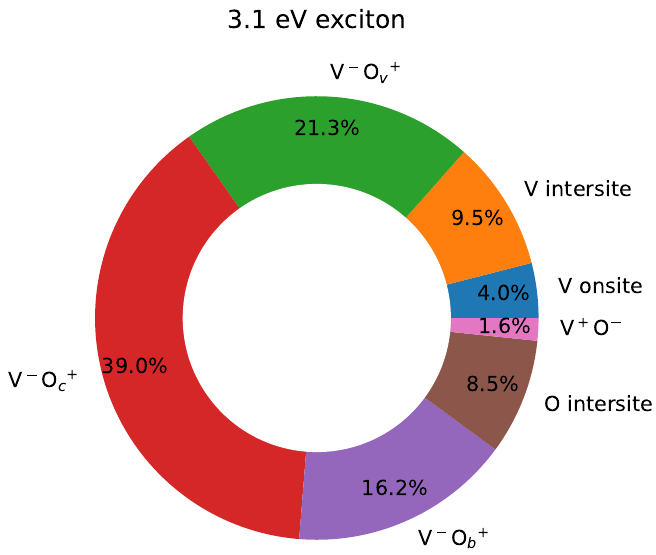}
  (f) \includegraphics[width=4.8cm]{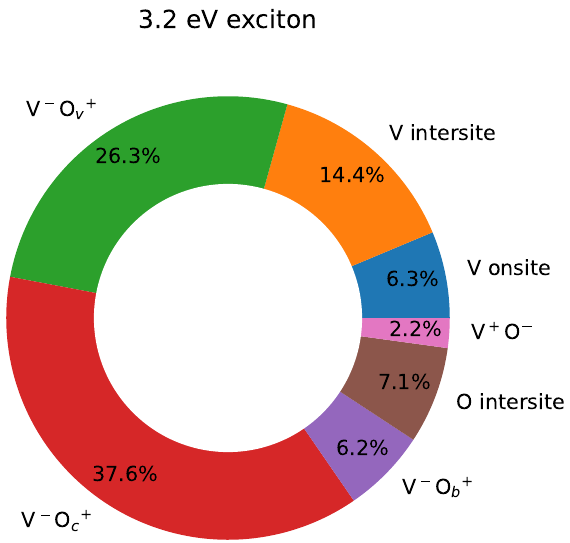}\\ 
  (g) \includegraphics[width=5cm]{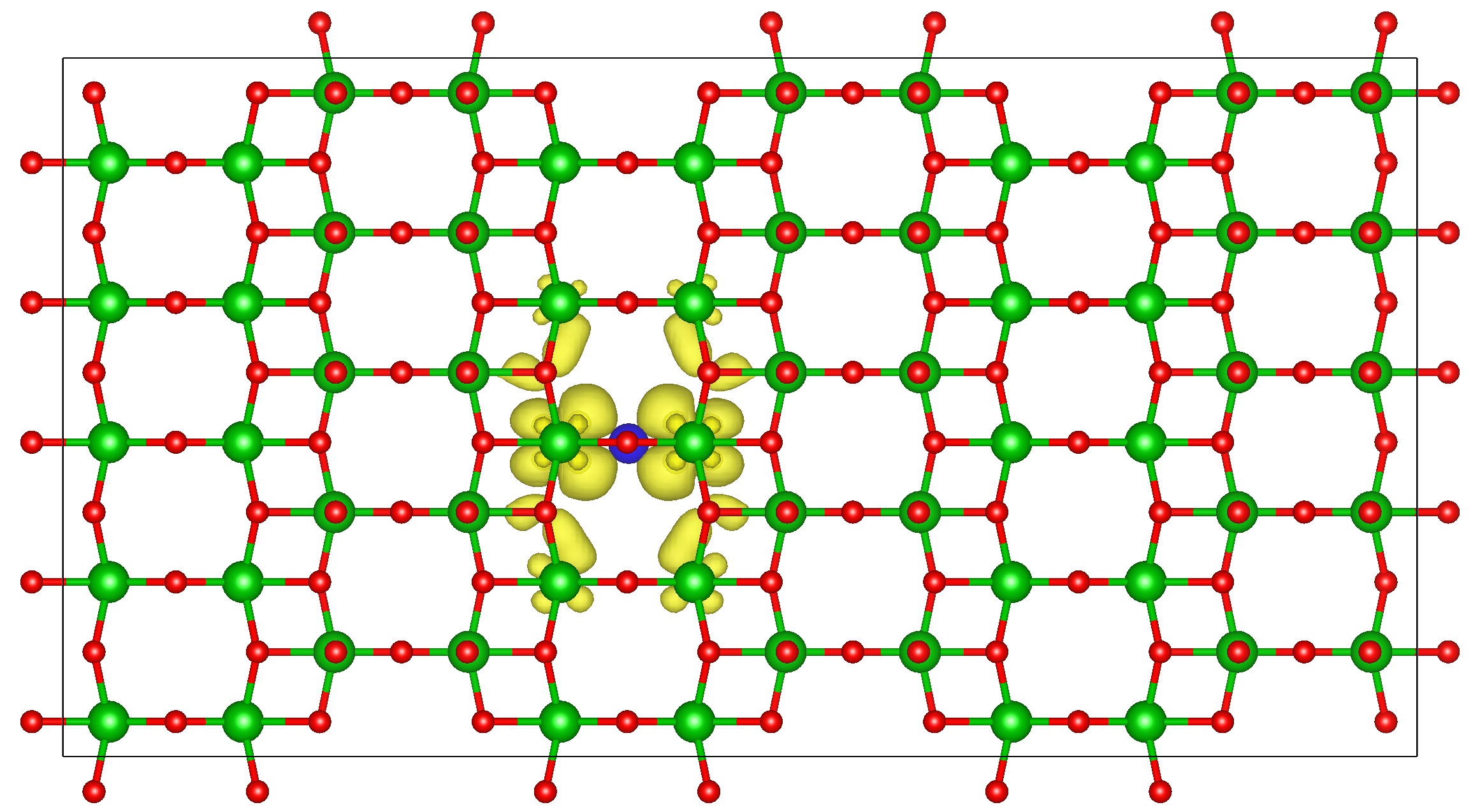}
  (h) \includegraphics[width=5cm]{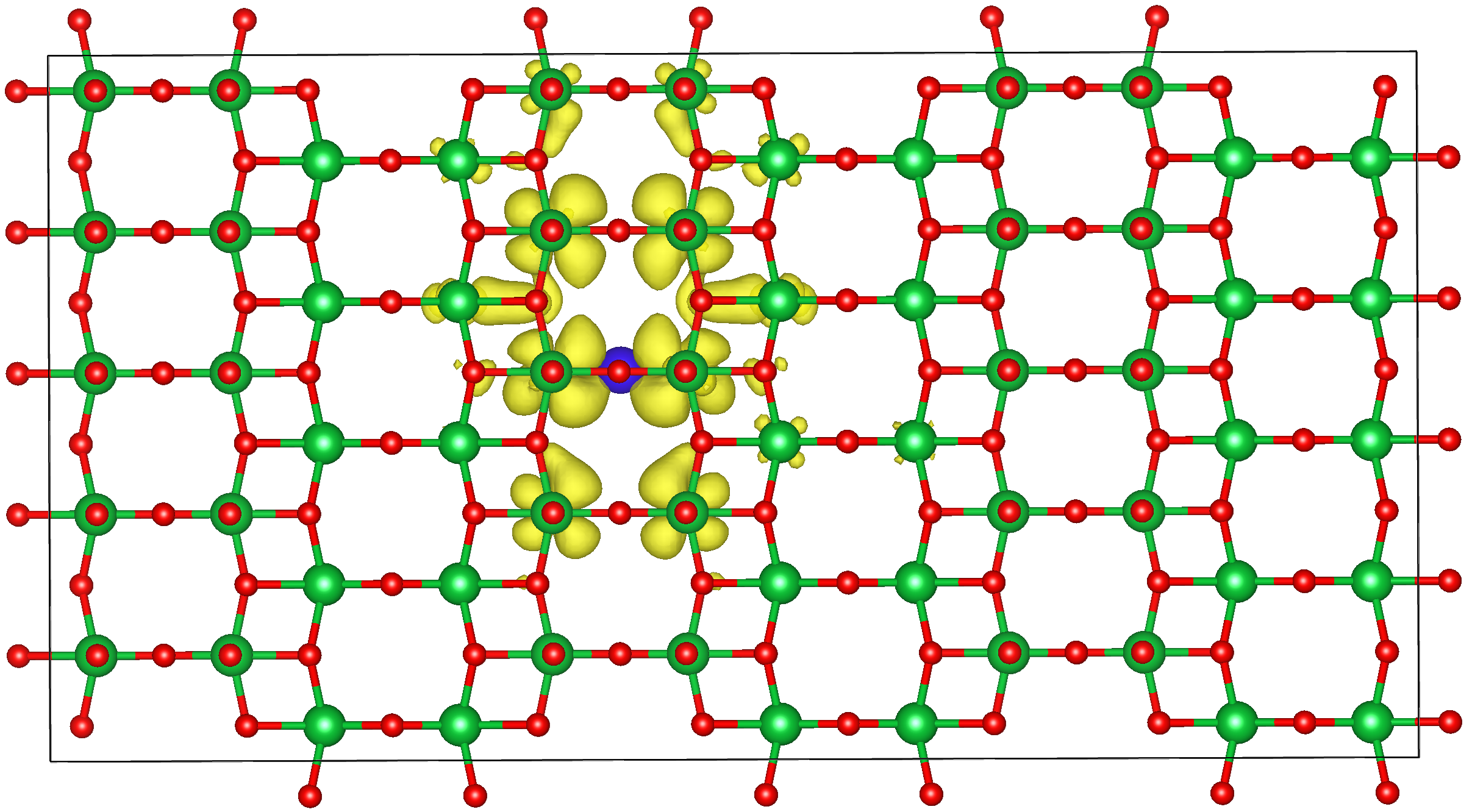}
  (i) \includegraphics[width=5cm]{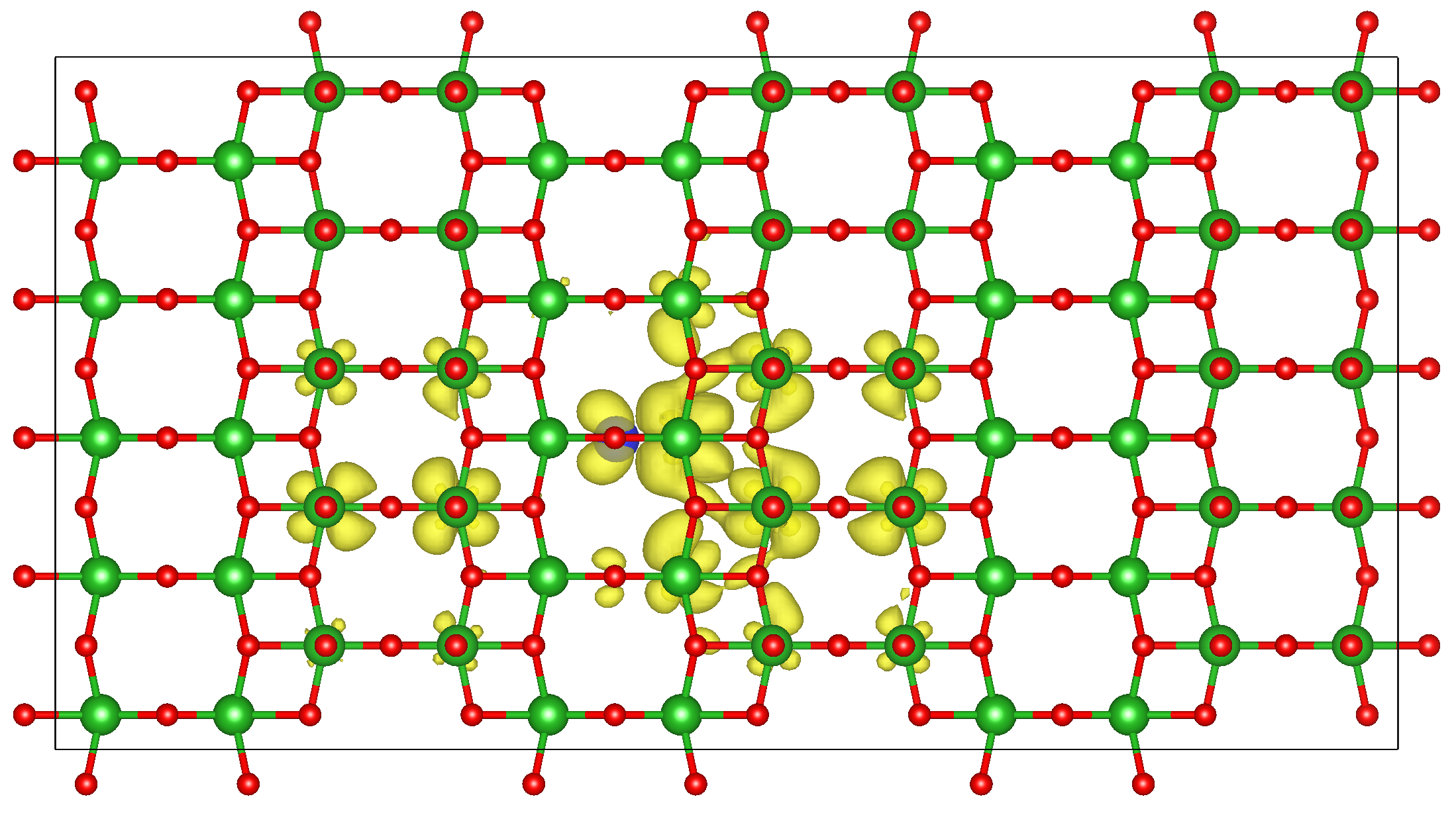}\\
  (j) \includegraphics[width=5cm]{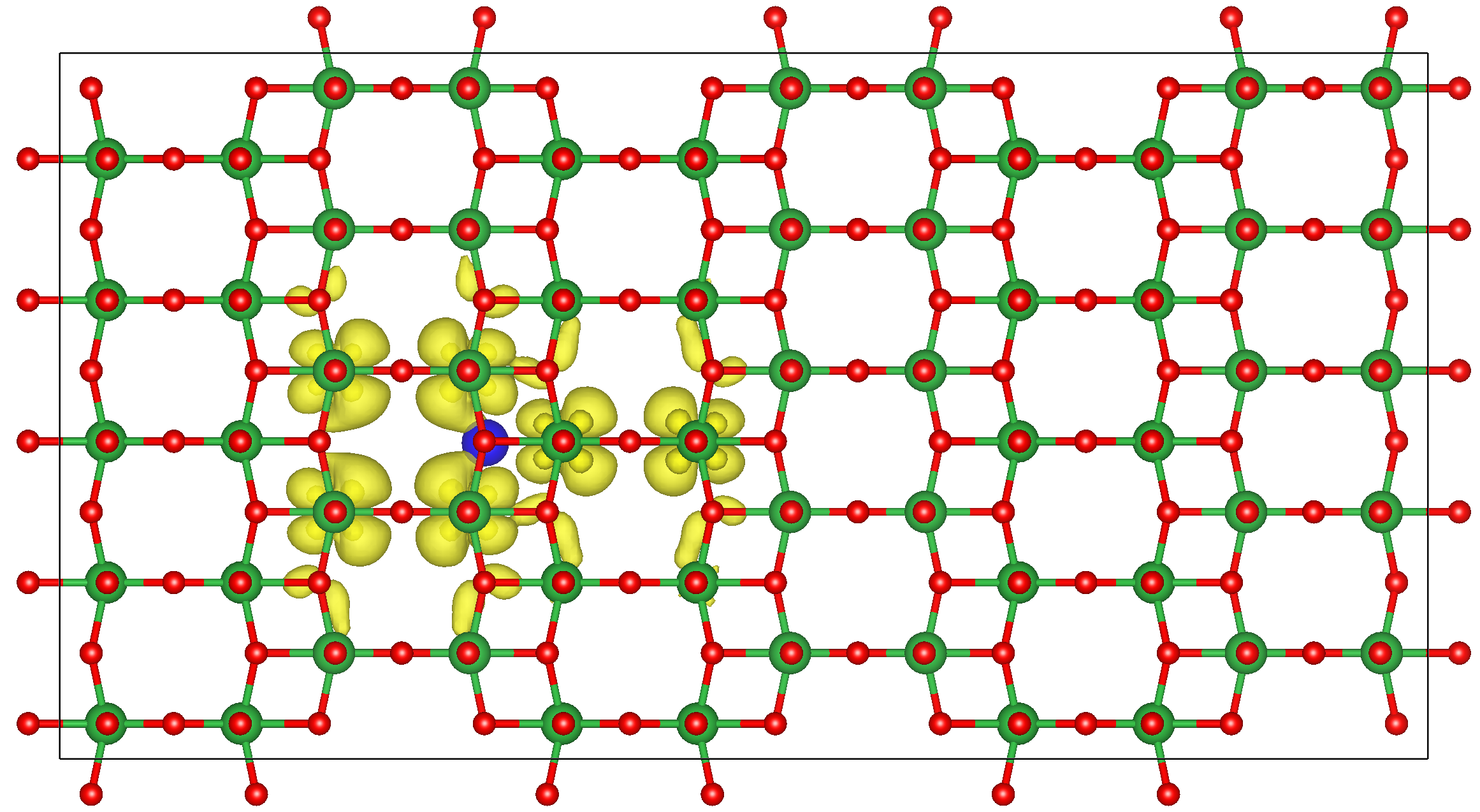}
  (k)  \includegraphics[width=5cm]{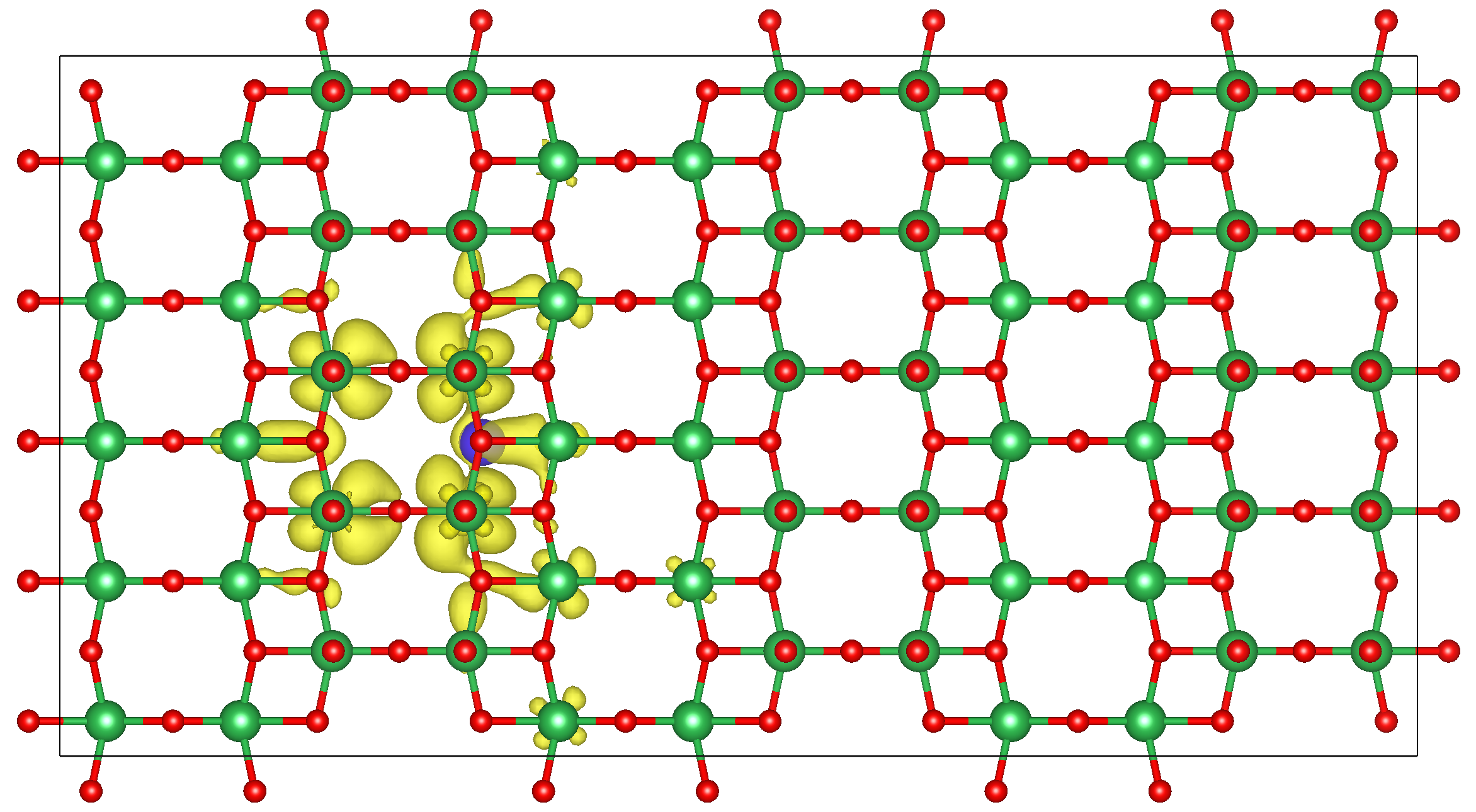}
  (l) \includegraphics[width=5cm]{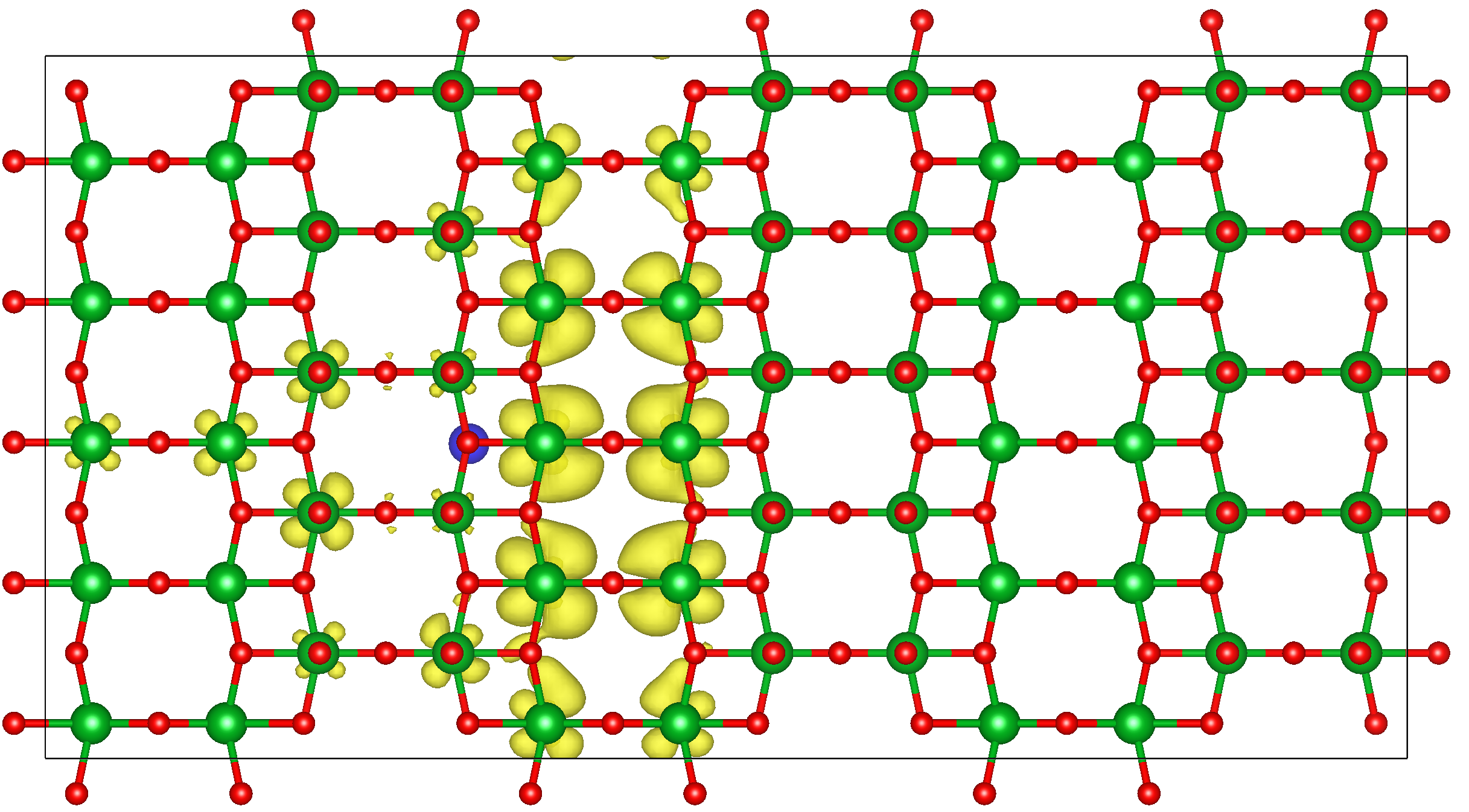}\\
  (m) \includegraphics[width=5cm]{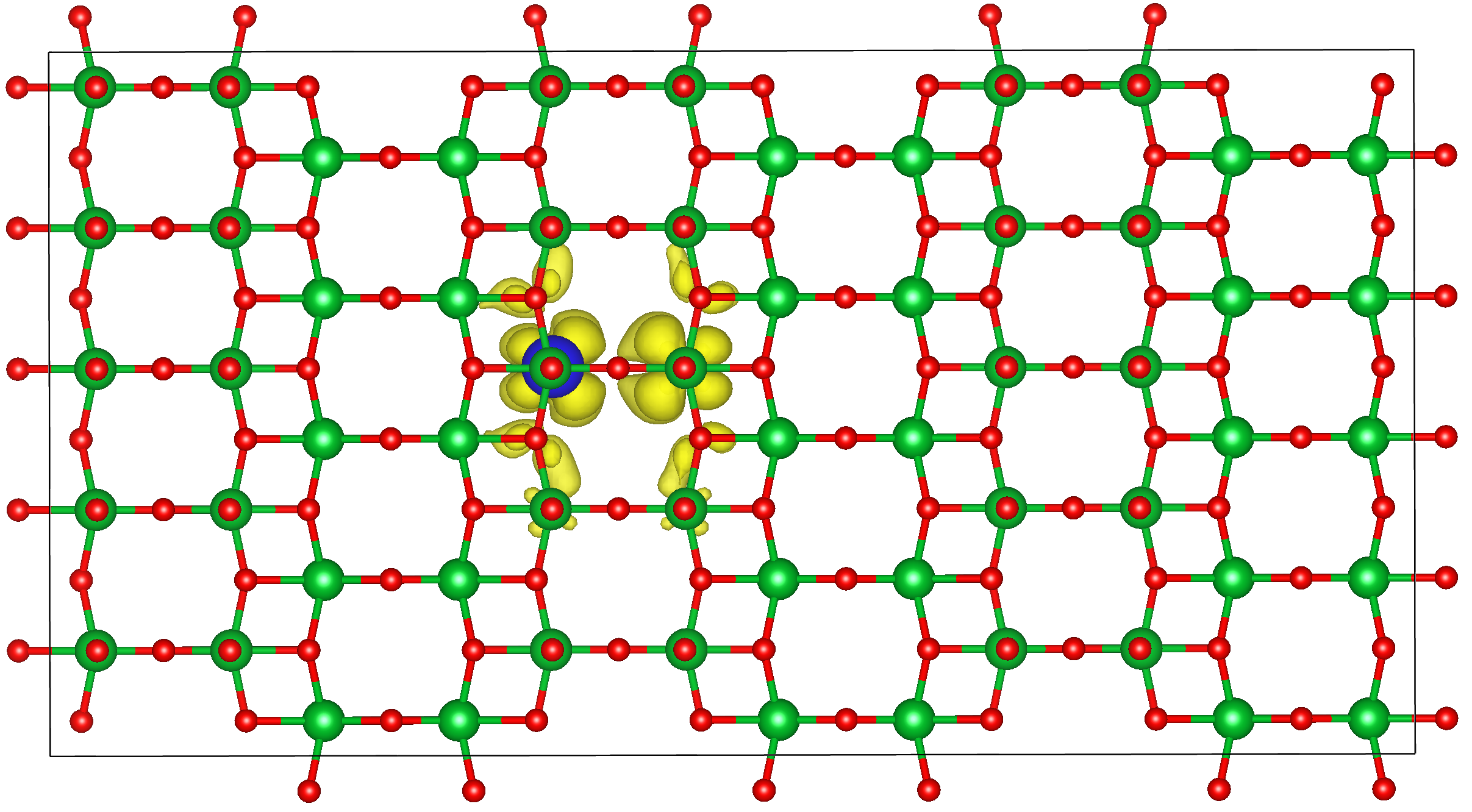}
  (n) \includegraphics[width=5cm]{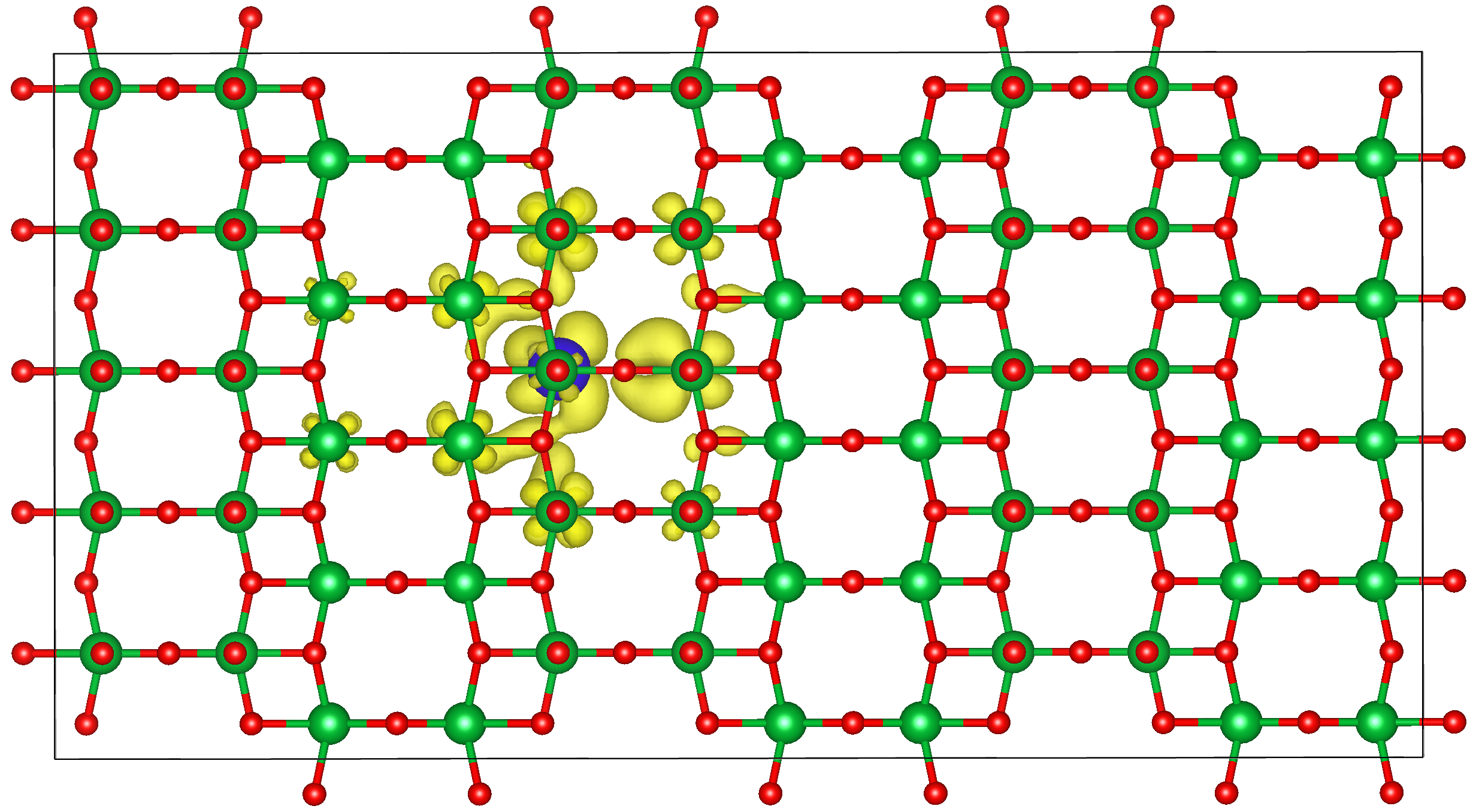}
  (o) \includegraphics[width=5cm]{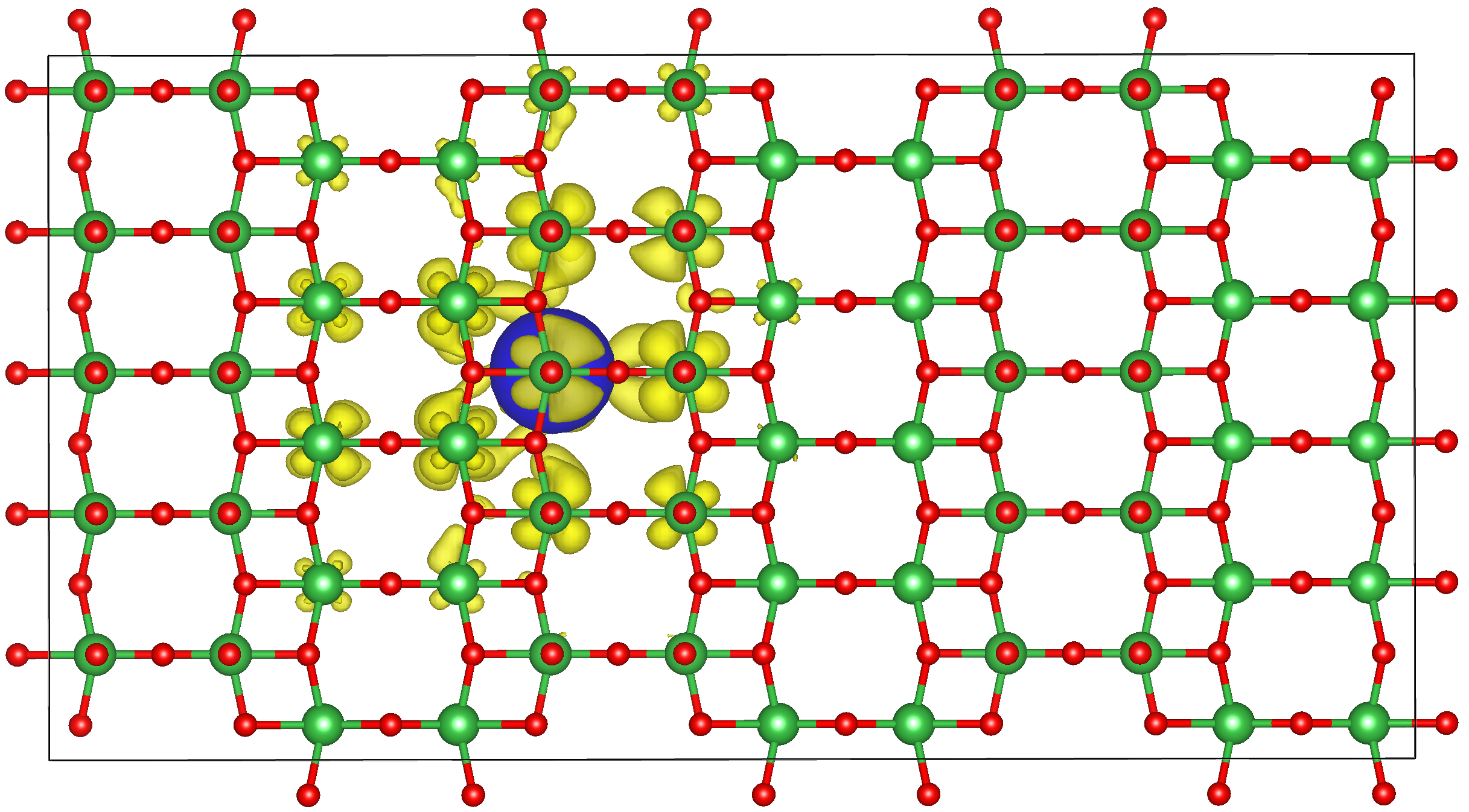}\\
  (p) \includegraphics[width=5cm]{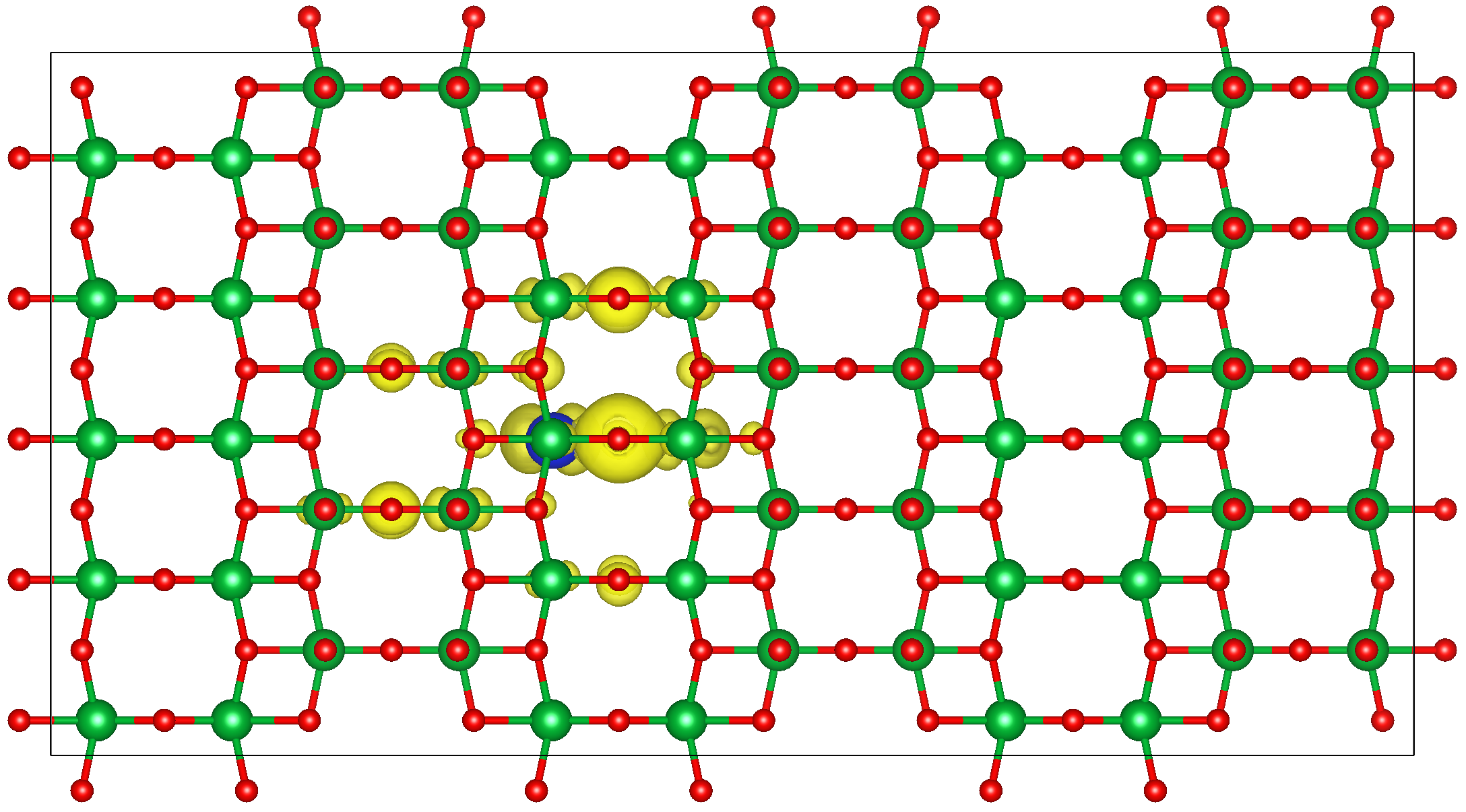}
  (q)\includegraphics[width=5cm]{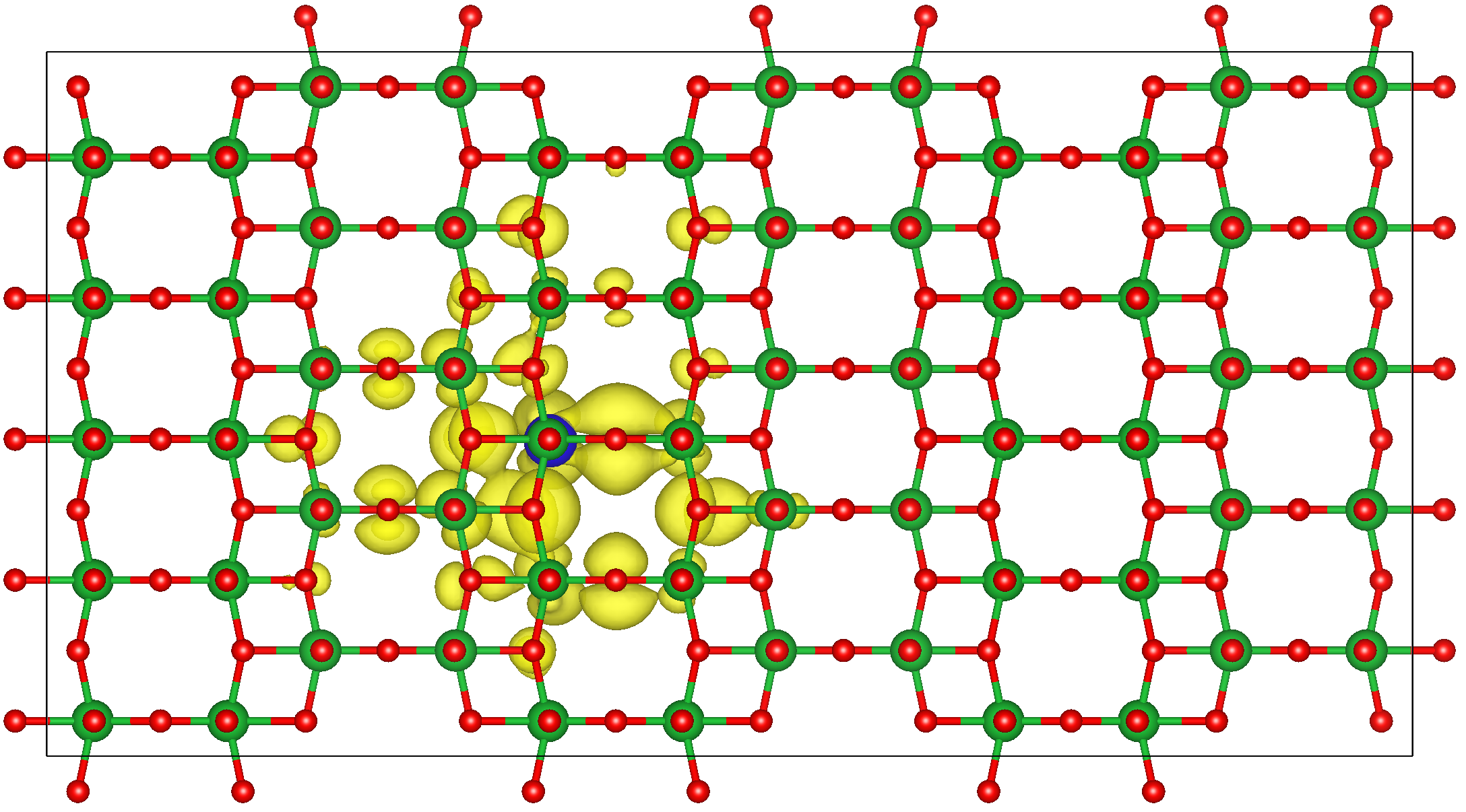}
  (r) \includegraphics[width=5cm]{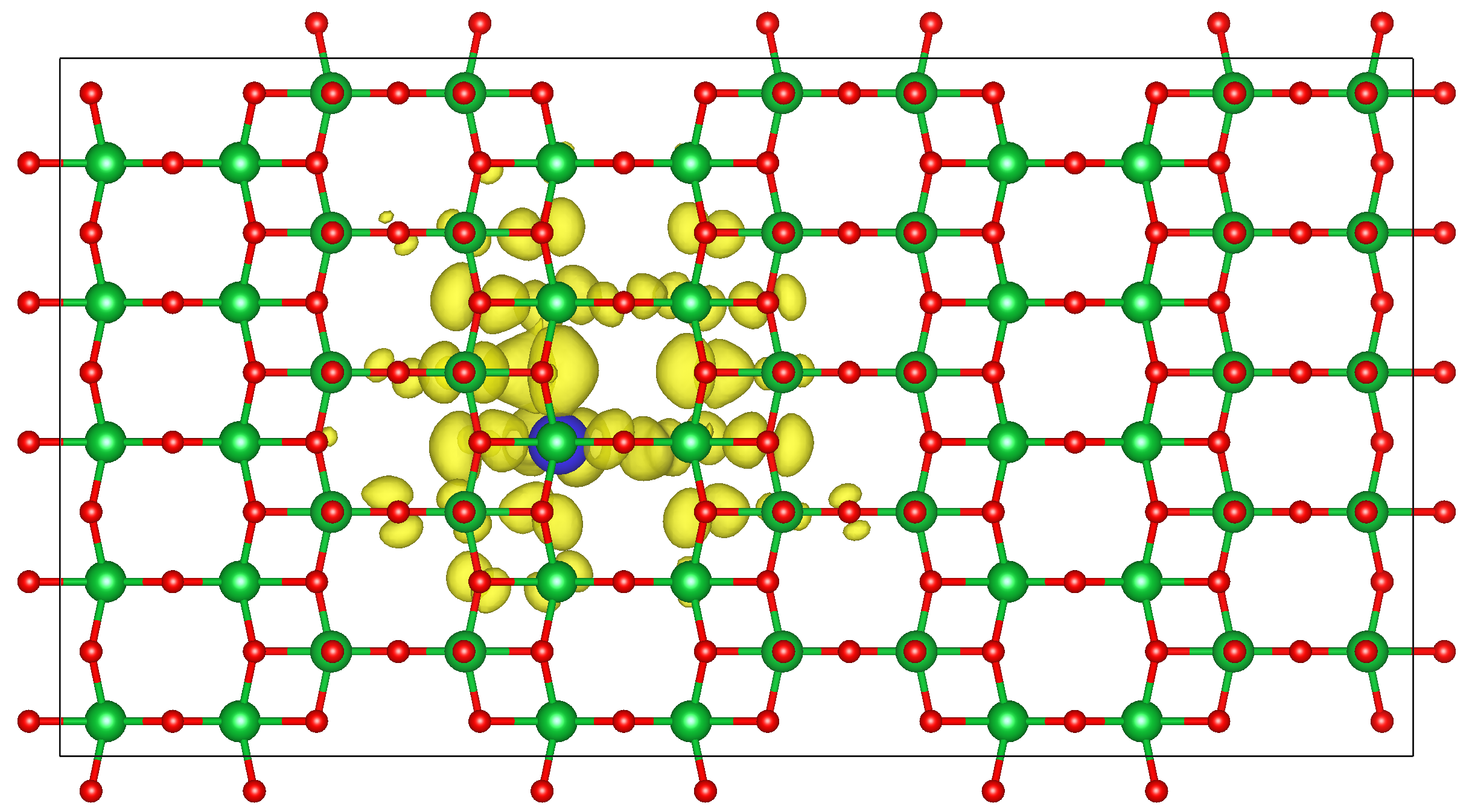}
  \caption{Exciton wavefunction
    analysis: (a-c) give the exciton weight $W^\lambda_{v(c){\bf k}}$
    along symmetry lines; (d-f) give integrated
    decompositions on atom pairs as a pie-chart; (g-i) give real space
    figures as function of ${\bf r}_e$ when hole is placed on the O$_b$,
    (j-l) on  O$_c$, (m-o) on O$_v$ and (p-r) as function ${\bf r}_h$ when electron is fixed at V. The location of the fixed hole or electron is indicated
    by the blue sphere. 
    Cases (a,d,g,j,m,p) refer to the dark 2.65 eV exciton,
    (b,e,h,k,n,q) to the 3.1 eV ${\bf E}\parallel{a}$ bright exciton and
    (c,f,i,l,o,r) for the 3.2 eV  ${\bf E}\parallel{\bf b}$ bright exciton.
         \label{figexpsi}}
\end{figure*}

We  study the composition of the excitons in various ways.
First, 
we show the bands that contribute significantly to a given exciton $\lambda$ by selecting a narrow energy window containing  just one exciton eigenvalue and by plotting
$W^\lambda_{v{\bf k}}=\sum_c |A^\lambda_{vc{\bf k}}|^2$ as a color weight on the band plot, where the
sum is over $c$, when plotting the weight on  the valence band $v$.
Similarly, $W^\lambda_{c{\bf k}}=\sum_v |A^\lambda_{vc{\bf k}}|^2$
gives the weight on the conduction bands.
These are shown in the first row of Fig. \ref{figexpsi}(a-c) for
different excitons of interest.
Next, in $\Psi^\lambda=\sum_{vc{\bf k}} A^\lambda_{vc{\bf k}}\psi^h_{v{\bf k}}\psi^e_{c{\bf k}}$, we can expand the Bloch functions $\psi_{v{\bf k}}$ into the muffin-tin orbital basis functions in a Mulliken analysis, and sum these over
angular momenta per atom to obtain a contribution per atom and hence per atom
pair of the exciton. The $\pm$ superscript indicates the hole or electron
atom location. This is a fully real space analysis. In other words, an inverse Fourier sum is applied to the LMTO basis Bloch functions depending on the
{\bf k}-mesh used. For a $N\times N\times N$ {\bf k}-mesh, we obtain contributions in a $N\times N\times N$ supercell in real space.
We select the most
important contributions and indicate them as a percentage on a pie-chart
in the second row (d-f).  For example V$^-$O$_c^+$ means all contributions
from an electron on a V and a hole on an O$_c$ regardless of the relative position of the two atoms. 
Third, we can pick a location for the hole and then display the probability
to find the electron around  it as a isosurface or fix the electron and
visualize the hole distribution. 
We here analyze the dark exciton at 2.65 eV, the ${\bf E}\parallel{\bf a}$ bright exciton at 3.1 eV and then the 3.2 eV ${\bf E}\parallel{\bf b}$ exciton
from left to right.

First, from the band weight plots, we can see that all three excitons
are derived primarily from the top valence band and lowest conduction
band with some smaller contributions from  bands farther away from the
band edges. They are very spread out in {\bf k}-space, and hence
localized in real space.  The localization in real space depends somewhat
on the arbitrary choice of isosurface value cut-off which we pick around 10 \%.
Nonetheless they are spread in real space over a few neighbor distances
in each direction.  One can see from the band plots that
the exciton weights are slightly different for the three excitons.
For example, the ${\bf E}\parallel{\bf a}$ exciton had a stronger
contribution from  $UTR$ and $SY\Gamma$  lines while the
${\bf E}\parallel{\bf b}$ exciton has larger contributions from $X\Gamma ZU$.
The dark exciton is even more equally spread in {\bf k}-space and hence even more localized
in real space. 
This is  consistent with a similar analysis by Gorelov\etal\cite{Gorelov22}.

The atom pair analysis shows that the exciton weights stem primarily
from electrons on V and holes on the various O. This is consistent
with the band analysis, since the lowest conduction bands are V-O antibonding
states and have primarily V-$3d$ content, while the top valence bands are V-O bonding
states and have primarily O-$2p$ content. It is interesting that the
different O do not contribute equally. For the dark exciton, the primary
contribution is from the O$_b$ $\sim$40\% with a small contribution
from chain oxygens  $\sim$9 \% and $\sim$30 \% of the vanadyl O.
This distribution occurs  in spite of the fact that each V has one O$_v$, three O$_c$ neighbors and only one O$_b$ is shared by two V across a bridge.
For the bright excitons, instead we see primarily
contribution around 40\% from the chain O and only  a small contribution (about 16 \% and 6 \% for ${\bf a}$ and ${\bf b}$ directions respectively)
from the bridge O and  and around 21--26 \% of the vanadyl oxygen.

We next show the real space figures for each exciton when the hole is
fixed on O$_b$, O$_c$ and O$_v$ and when the electron is fixed on V.
We can compare the 3.1 and 2.65 eV excitons for the hole fixed on the bridge
O with the work of Gorelov \etal\cite{Gorelov22}.
In that paper the exciton appeared more extended in the {\bf a} direction
perpendicular to the chains,  and a simple tight-binding model with exciton
wave functions centered on the V-O$_b$-V bridge
was developed  to understand their  spread,
comparing in particular the dark and the bright exciton
for ${\bf E}\parallel{\bf a}$ as even and odd partners to each other in
their ${\bf k}$ and $-{\bf k}$ components. 
In retrospect that model, while instructive,  may be somewhat oversimplified.\cite{Gorelovnote,Gorelov24}.

One might ask to what extent 
the real space distributions are sensitive to the precise location of the fixed hole (or electron). The code used to make these figures
snaps the position we give as input 
to the nearest grid point in the real space mesh and this can sometimes
be slightly off from the more symmetric atom position we target. For example,
Fig. \ref{figexpsi}(i) appears to have the electron distribution skewed to the right of the bridge O. Nonetheless in (g) and (h) we choose the exact same
hole location and yet these appear more symmetric. On the other hand, in
(k) and (l) we use the same O$_c$ position and yet for the 3.1 eV exciton the
wave function spreads more the the left and for  the 3.2 eV one more to the right. In view of their different {\bf k}-space localization, these appear to be genuine differences between these excitons and not just artifacts of the precise
location of the fixed particle in the exciton and we further tested that
they are robust to small displacements of the assumed hole position.  
Complementary information is gained by fixing the electron on a V and examining the corresponding hole distribution. These are show in part (p-r) of Fig. \ref{figexpsi}
In these figures we can recognize the O-$p$ like character, while in the
previous ones, we can recognize the $d_{xy}$ like character on V.

The consistent picture that emerges from these various visualizations is that
the dark exciton at 2.65 eV is significantly more localized than the
two bright excitons considered here. They have a rather complex distribution
spread over a size of about 5-15 \AA\  and are charge transfer like excitons.
Overall, these examples confirm the main finding from Gorelov \etal\cite{Gorelov22} that the excitons
are not Frenkel excitons, which one might  expect to stay localized on a
single atom or molecular fragment like the V-O$_b$-V bridge, but are more
spread out
than one would expect for such
large exciton binding energies. However, they are  more complex than previously thought. Similar strongly anisotropic (almost unidirectional) and strongly bound excitons have been observed in the puckered two-dimensional magnet CrBrS~\cite{klein2023bulk,shao2024exciton}. 
Much as the strongly bound excitons in V$_{2}$O$_{5}$ those excitons also extend up to $\sim$3-4 unit cells. However, in strong contrast to the excitons in V$_{2}$O$_{5}$, excitons in CrBrS originate from partially filled $d$-states and are magnetic in nature and have both large on-site $dd$ and significant inter-site dipole $dp$ characters~\cite{ruta2023hyperbolic} to them. The V$_{2}$O$_{5}$ excitons, however, have barely any onsite components and mostly share the electrons and holes on the V-O ($dp$) dipole. In other words, as already pointed out by Gorelov \etal\cite{Gorelov22} they
can be viewed as charge-transfer excitons. 
It is in that sense that these excitons are significantly different from $dd$ Frenkel excitons as observed in several strongly correlated ferro- and anti-ferromagnets\cite{acharya2022real,acharya_theory_2023}.

\subsection{Monolayer}
\subsubsection{Quasiparticle and optical gaps}
\begin{figure}
  \includegraphics[width=8cm]{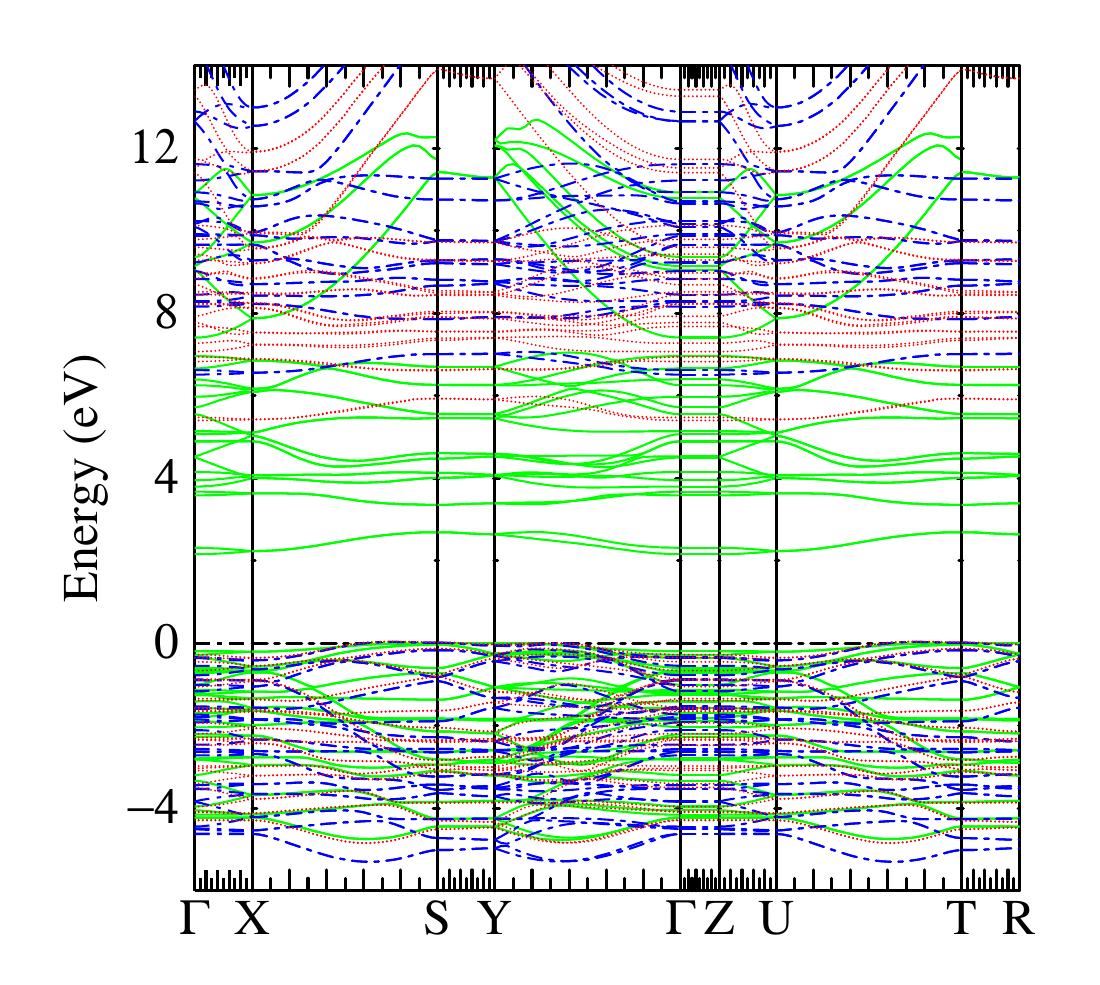}
    \caption{Band structure of monolayer V$_2$O$_5$ in GGA (green solid line) , QS$GW$ (blue dot-dashed)  and QS$G\hat{W}$ (red-dotted) for $c/a=1.5$. \label{fig:band-mono}
   }
\end{figure}

\begin{figure}
  \includegraphics[width=9cm]{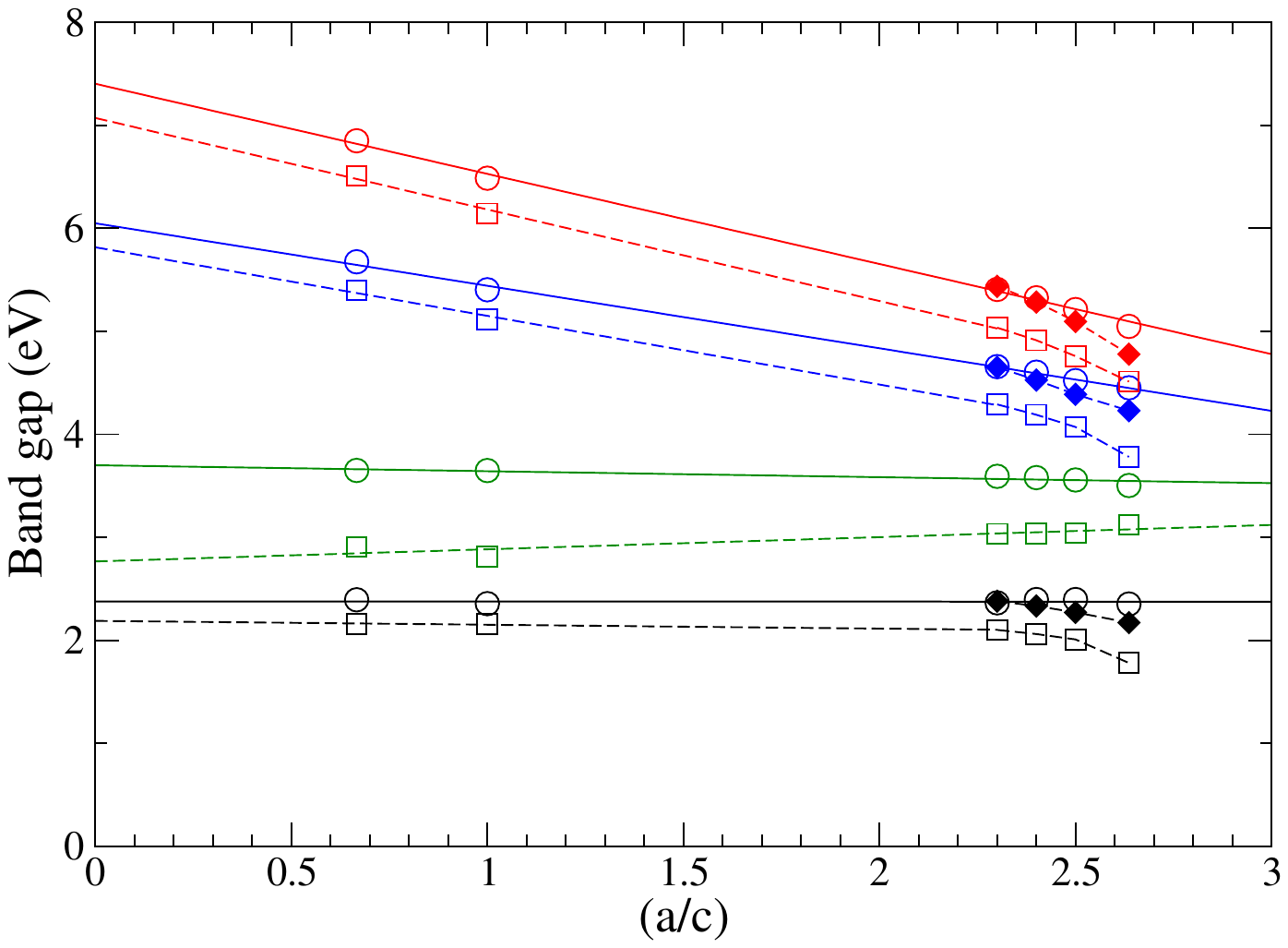}
  \caption{Quasiparticle gaps  in V$_2$O$_5$: direct gap at $\Gamma$  (circles)  lowest direct gap (diamonds),
    and indirect gap (squares),  as function of $a/c$, in GGA (black), QS$GW$ (red) and
    QS$G\hat{W}$ (blue) with straight line interpolations in the linear region; 
     lowest exciton gap  for ${\bf E}\parallel {\bf a}$ in BSE (green) using either QS$GW$ (solid line), 
    or QS$G\hat{W}$ self-energy (dashed line).   
    \label{fig:gapsmono}}
 \end{figure}

Having established good agreement with prior work for V$_2$O$_5$ in spite of some  differences, we move on to
study the monolayer.  To calculate the monolayer, we simply increase the distance between the V$_2$O$_5$ layers
by increasing the $c$-lattice constant and keeping the layer atomic positions fixed.
Using the  $z$ coordinate difference between the vanadyl oxygens sticking out on either side of the layer
as a measure of the thickness of the layer, the layer has a thickness of 4.096 \AA.  The c lattice constant
is 4.368 \AA\  and the V-O$_v$ vertical distance is 1.575 \AA, so between the O$_v$ of one layer and
the V above it in the next layer, the distance is 2.793 \AA. When we set the $c_{mono}=a_{bulk}$
the vacuum thickness is 7.416 \AA\ and the distance from the O$_v$ to the next layer  V is 9.9\AA.
Using $c_{mono}=1.5a_{bulk}$ the vacuum layer is 13.17\AA\  and the vertical distance from the O$_v$ to the V above it is 15.69 \AA.  These seem sufficiently large to represent well isolated monolayers from the point of view of having negligible hopping between the layers. In fact, we will show that by this distance the GGA gap is  well converged but the
QS$GW$ gap is not. The band structure of the monolayer
using  $c_{mono}=1.5a_{bulk}$ is shown in Fig. \ref{fig:band-mono}. Note that strictly speaking for a monolayer with infinite
separation, 
the Brillouin zone edge in the $c$-direction $k_z=\pi/c$ should go to zero. However, by showing the bands also in the $k_z=\pi/c$ plane, $ZUTR$,  we show explicitly how flat the bands are along the $c$ direction for the $c/a$ used.
The bands in the $ZUTR$ plane are then equal to the $\Gamma XSY$ and the extent to which this is true indicates
whether the $c$ distance is large enough to avoid inter-layer hopping. 
To check 
the convergence of the gaps we plot the direct and indirect band gaps 
as function of $a/c$ in Fig. \ref{fig:gapsmono} and use a linear extrapolation
of the QS$GW$ gaps in the range where the behavior does indeed become linear.

\begin{table}
  \label{tabindirectgapmono}
\end{table}

The band structure plot Fig. \ref{fig:band-mono} shows that already at the
GGA level, the indirect gap is slightly increased compared to the bulk, primarily  because the highest valence band in the $TRUZ$ plane (at $k_z=\pi/c$) is now almost the same as in the $\Gamma XSY$ plane (at $k_z=0$). The upward dispersion from $\Gamma-Z$ in the bulk case is missing.  This indicates that this dispersion
is related to the interlayer hopping  interaction in the bulk.
Several  changes happen in the band structure: the smallest direct gap, which in bulk
occurs at  $Z$ now occurs at $\Gamma$ because the bands along the $\Gamma-Z$ direction become flat. Secondly, the indirect gap, which in bulk occurs between
the VBM at or near $T$ and the CBM at $\Gamma$  now shifts to a point between X-S and $\Gamma$.
The self-energy shifts  are significantly higher than in the bulk. 

In Fig. \ref{fig:gapsmono}, we can see that
these changes occur as soon as the layers become decoupled already for a modest increase in interlayer distance
($a/c$=2.3 or $c$=5 \AA). The smallest direct gap becomes equal to the direct gap at $\Gamma$ and
in the GGA, the band gaps have essentially converged at this point and stay constant. 
On the other hand, the QS$GW$ and QS$G\hat{W}$ gaps  keep on increasing linearly as we further increase $c$. 
This slow convergence with the size of the vacuum region  is caused by the long-range
nature of the self-energy $\Sigma$ which is proportional to the screened Coulomb interaction $1/\varepsilon r$ because of
the screened exchange term. With increasing size of the vacuum the effective dielectric constant of the system
becomes smaller. Effectively, the long-range part of the Coulomb interaction becomes unscreened and dominated
by the vacuum or surrounding medium for a thin 2D system.
This is well known since the work of Keldysh
\cite{Keldysh79} and discussed in detail in Cudazzo \etal \cite{Cudazzo11}.
The monolayer is thus predicted to have a significantly higher gap than bulk layered V$_2$O$_5$.
The linearly extrapolated direct quasiparticle gaps are 7.4 eV  and 6.1 eV in the QS$GW$ and QS$G\hat{W}$ approximations.
The difference between direct and indirect gap stays approximately constant as we increase $c$. Also, the difference between QS$GW$ and QS$G\hat{W}$ stays
more or less constant.
 
On the other hand, the BSE optical gap stays almost constant. The lowest optical gap shown in Fig. \ref{fig:gapsmono}
is  for ${\bf E}\parallel{\bf a}$ and is a mixture of various direct interband transitions spread throughout {\bf k}-space. It is not dominated by the lowest gap  direct gap (which is at $Z$ in the bulk case) as we have seen in
Fig. \ref{figexpsi}. It does not show the initial increase of the direct and indirect gaps
as we start increasing $c$. It also does not increase as the QS$GW$ gaps. This is because the exciton binding energy
is also proportional to $W$ and hence an increase in $W$ due to lower screening results both in
an increased self-energy and quasiparticle gap  but is compensated by an increased exciton binding energy.
The optical gap is thus expected to change only minimally. This applies both
when we use $W$ or $\hat{W}$.
In the latter case, the gap seems to go slightly down for larger $c$, but this is within the error bar. It might indicate that the increase in $\hat W$
with $c/a$ is more directly reflected in the exciton
binding energy than in the quasiparticle self-energy. 
 This also implies  that the exciton gap will be less affected by substrate dielectric screening if the monolayer is placed on top of  a substrate.
 Our optical calculations only provide the excitons derived from direct transitions and the bound excitons are  a mixture of
 vertical band-to-band transitions spread over the whole Brillouin zone. Therefore,  the lowest exciton gap (green lines)
 does not show the rapid increase with  increasing $c/a$ starting from the bulk values.   
 There should also be an indirect exciton related to the indirect excitation of an electron-hole pair via a combined photon and phonon interaction in second order perturbation theory and modified by the electron-hole interaction.  At present
 we cannot calculate this but expect it to follow the dependence of the indirect quasiparticle gap as function of
 interlayer distance. In fact, this indirect gap changes {\bf k}-point location  and thus the character of the
 corresponding exciton will also change.  Assuming that the exciton binding energies of direct and indirect excitons
 are similar we note that the indirect quasiparticle gap between bulk and monolayer changes by about 0.1 eV
 and hence we expect a similar change in indirect exciton with interlayer distance.

\begin{figure}
  \includegraphics[width=9.5cm]{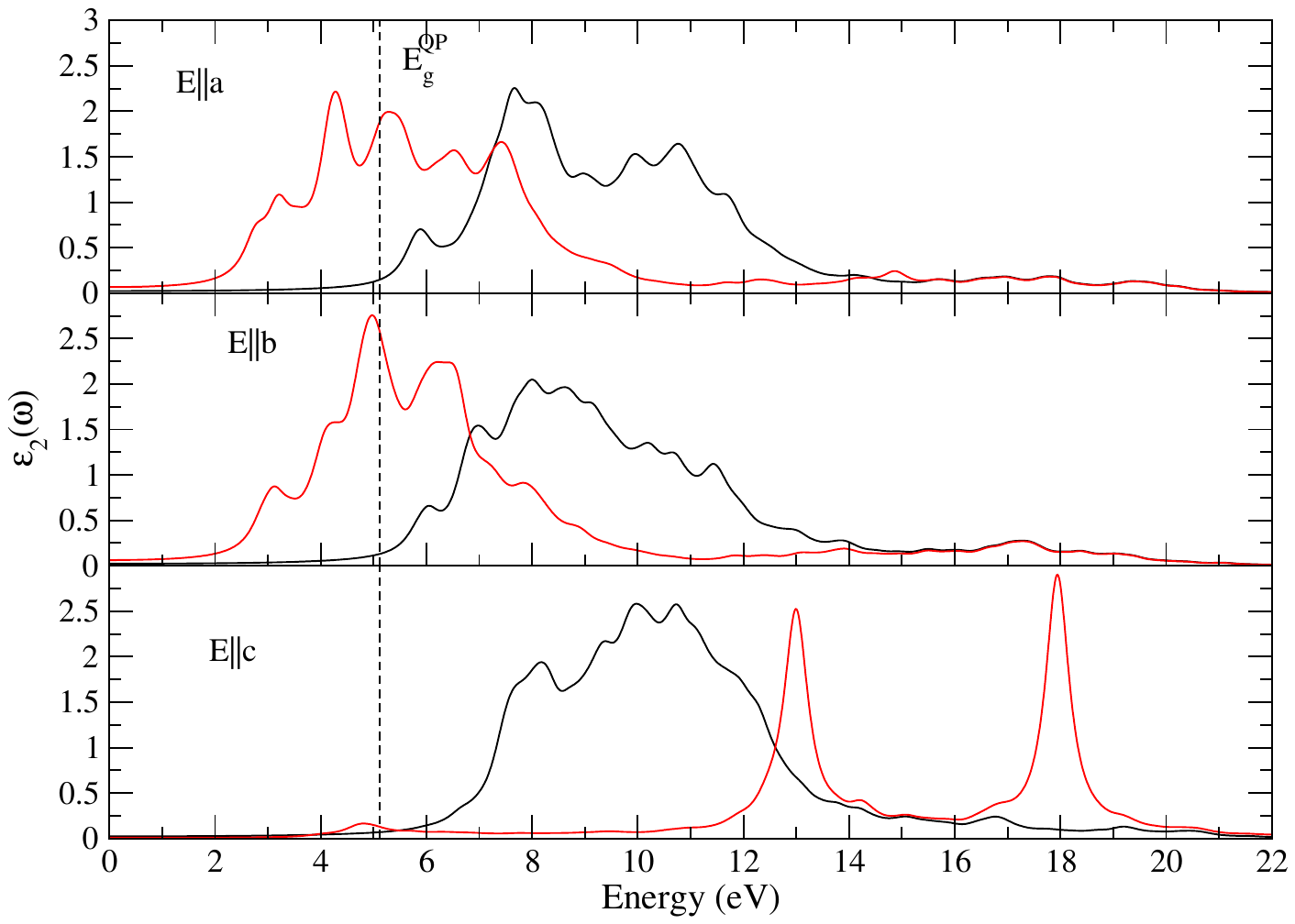}
  \caption{Imaginary part of the dielectric function in IPA and BSE for the monolayer limit  $c/a=1.5$ \label{figepsml}}
\end{figure}

\begin{figure}
  \includegraphics[width=9.0cm]{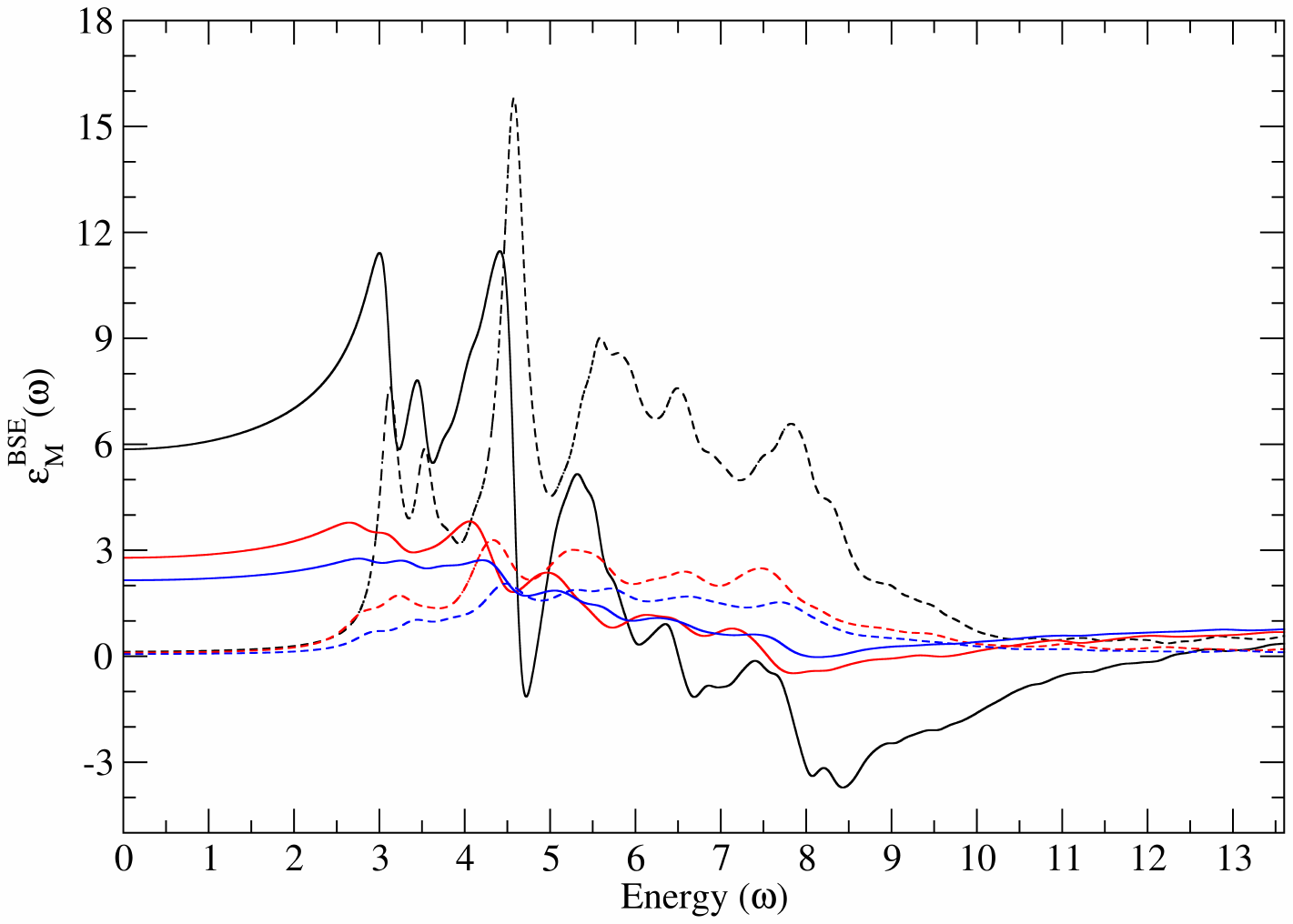}
  \caption{Real and imaginary part of the dielectric function for polarization
    ${\bf E}\parallel{\bf a}$ as function of interlayer spacing;
    solid lines: $\varepsilon_1$, dashed lines $\varepsilon_2$; bulk (black), $c/a=1$ (red) , $c/a=1.5$  (blue).\label{figlayer}}
 \end{figure}

Next we examine the  dielectric functions of the monolayer representative cells as function of interlayer distance  in Figs. \ref{figepsml} and \ref{figlayer} in some more detail. First, in Fig. \ref{figepsml} 
we can now see an even stronger suppression of the $\varepsilon_2(\omega)$
in the BSE for ${\bf E}||{\bf c}$.
Again, at higher energies, sharp features occur for the
polarization perpendicular to the layers but these are dismissed as
unrealistic artifacts from the BSE active space truncation.
This indicates that the local field effects are even stronger in the monolayer case.  
The excitons are still prominent for the in-plane polarizations, but the lowest peaks still occur near 3.0 eV not too far from the bulk case. Still, the shape of the $\varepsilon_2(\omega)$, \ie the exciton spectrum,
is significantly different from the bulk case.

Next, we look a little more closely at the change in dielectric functions, both real and imaginary parts, 
as function of interlayer spacing in Fig. \ref{figlayer}.
We can see first of all that the amplitude of the $\varepsilon_2(\omega)$
and the values of $\varepsilon_1(\omega=0)$ are much reduced in the
monolayer cases compared to the bulk and increasingly more so as the
thickness of the vacuum layer increases.   This can simply be understood in terms of  a model of capacitors in series. Essentially, there is a thicker and
thicker region of relative dielectric constant $1$ in between the layers. Since the capacitance is
inversely proportional to its thickness $d$ and proportional to the dielectric
constant in that region, the effective dielectric constant can be obtained
from adding the capacitance of the layer and of the vacuum region in series,
which gives
\begin{equation}
  \frac{c}{\varepsilon_{eff}}=\frac{c_b}{\varepsilon}+\frac{c-c_b}{1}
\end{equation}
where $c_b$ is the $c$-lattice constant for bulk and $c$ the one in the monolayer model.  In the limit $c\rightarrow\infty$ this goes to 1, the dielectric
constant of vacuum,  and in the limit $c\rightarrow c_b$ it gives $\varepsilon_{eff}=\varepsilon$ of bulk V$_2$O$_5$
We caution that these dielectric functions of the periodically repeated layers do not represent the true
dielectric screening behavior inside an isolated monolayer but rather that of the overall system including vacuum.
For $c\rightarrow\infty$ the overall dielectric function would go to $1$ as it becomes dominated by vacuum.  On the other hand the screening in two dimensions becomes strongly distance dependent with qualitatively different behavior
at distances smaller than the thickness of the layer and larger than it, as is well known since the work of Keldysh\cite{Keldysh79} and  Cudazzo \etal\cite{Cudazzo11}.

\begin{figure}
  \includegraphics[width=8.5cm]{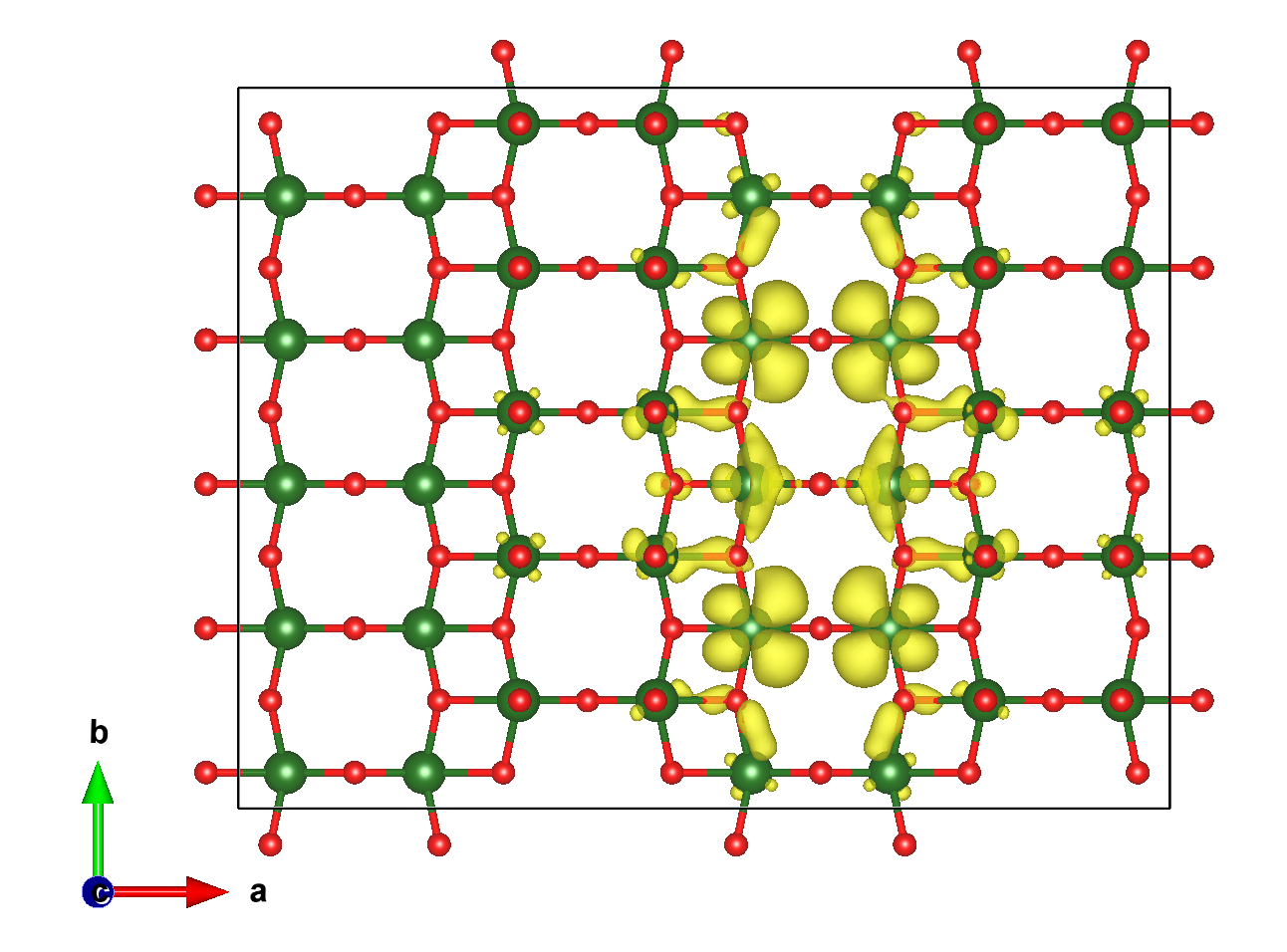}
  \caption{First bright exciton ${\bf E}||{\bf a}$ in monolayer V$_2$O$_5$ for the hole place on the bridge oxygen
    as function of electron position and calculated for $c/a=1.5$.
    \label{figexbrmono}}
\end{figure}

\begin{figure}
  \includegraphics[width=8cm]{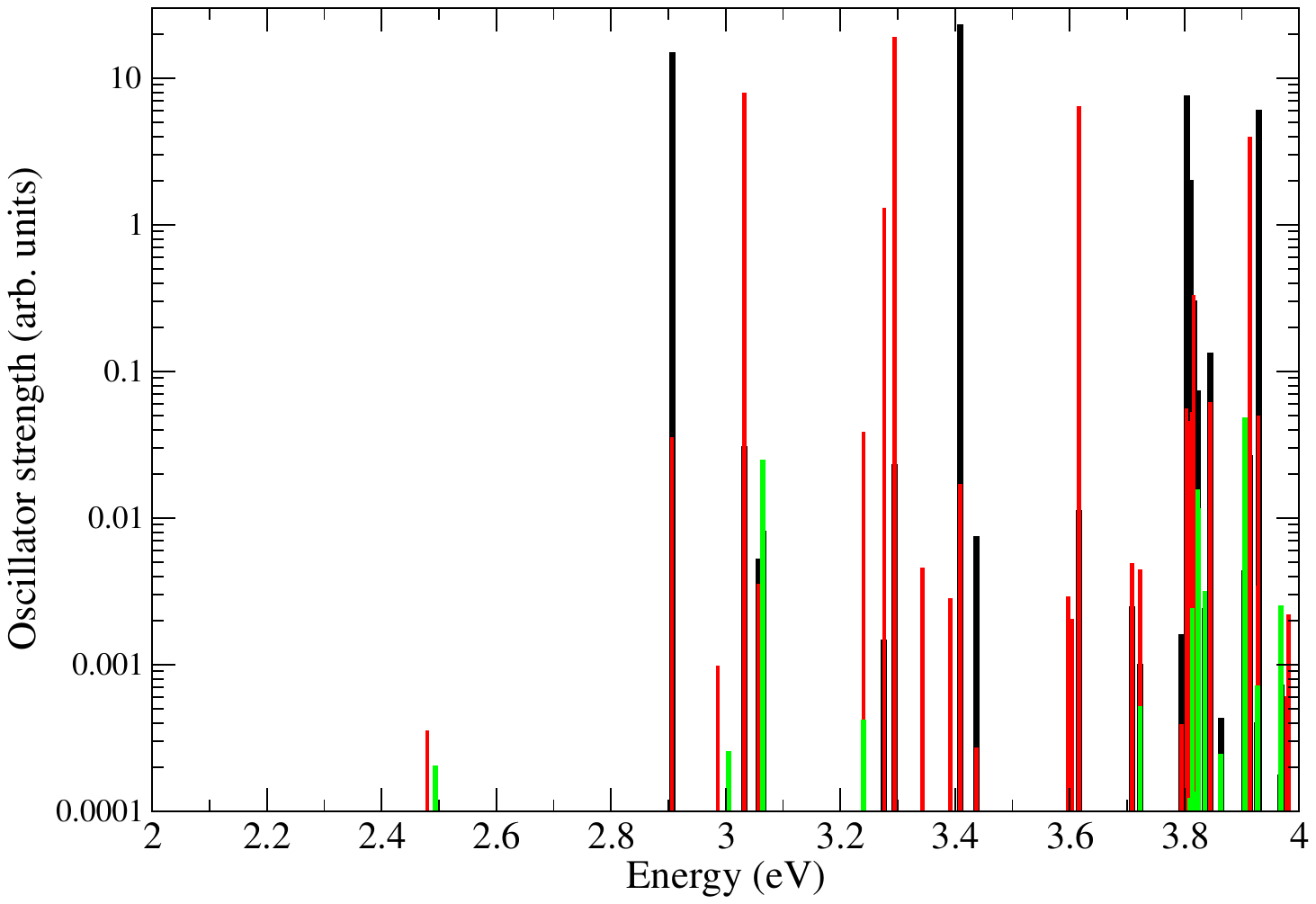}
  \caption{Relative oscillator strengths of exciton eigenvalues for the monolayer case $c/a=1.5$ on a log scale: black , red, green correspond to $a$, $b$, $c$ polarizations. \label{figmonolog}}
\end{figure}

The real space spread of the first bright exciton for ${\bf E}\parallel{\bf a}$
in the monolayer is shown in Fig. \ref{figexbrmono} for the hole placed on the bridge O$_b$ and as function of electron position.   It looks quite similar to that in the bulk case, shown in Fig. \ref{figexpsi}h. The difference is it that is entirely confined to
one monolayer while in the bulk case, it spreads slightly to neighboring layers.
This is not visible in the projection figures here but can be ascertained by viewing the exciton wavefunction from different angles. Its spread in ${\bf a}$ direction  appears slightly
larger here than  in Fig. \ref{figexpsi}h  but this is because we here used
a $2\times5\times1$ mesh. Apparently two {\bf k}-points in the {\bf a}
direction is not yet sufficient to completely avoid overlap of the excitons in adjacent
cells from the periodic boundary conditions  in the {\bf a} direction.
We note that similar to the bulk, dark excitons also exist at lower energy for the monolayer.
In Fig. \ref{figmonolog} we show the oscillator strengths of the  exciton eigenvalues
up to 4 eV on a log scale. Similar to the bulk case, we find a  pair of dark excitons
near  2.48, 2.49 eV, while the first bright excitons occur at 2.91 and 3.03 eV for ${\bf E}\parallel{\bf a}$ and ${\bf E}\parallel{\bf b}$ respectively.  Surprisingly, these lie actually slightly lower than the bulk case, even though the quasiparticle gap was strongly increased.
This is consistent with the slight decrease for increasing $c$ seen in Fig. \ref{fig:gapsmono} for the QS$G\hat W$ case.  

\subsubsection{Comparison to experiment}
Monolayer V$_2$O$_5$ has not yet been realized although attempts at exfoliation have resulted in ultrathin layers of order 8-10 atomic layers thick.\cite{Sucharitakul17} Only recently, layers as thin as bi- or trilayer of V$_2$O$_5$ were realized by sonification after swelling of the interlayer distance by intercalation with formamide molecules as reported by Reshma \etal\cite{Reshma21}.
These studies showed an increase in optical absorption edge by about 1.3 eV for the thinnest samples which
contained individual layers of order 1.1-1.5 nm, corresponding to 2-3 layers.
It is not straightforward to interpret the onset of the Tauc plot as the direct gap because of the
large excitonic effects and disorder related band tailing effects.
The Tauc plot prediction of an absorption coefficient proportional to
$\sqrt{E-E_{gap}}$ for direct allowed transitions is valid only for band to band transitions. However, including polaritonic
effects it may also correspond to indirect excitons \cite{Elliott57}.
The value reported for bulk in \cite{Reshma21}  is 2.39 eV, which  is close to the
gap reported by Kenny and Kannewurf\cite{Kenny66} but much smaller than the direct excitonic peak
seen in spectroscopic ellipsometry \cite{Gorelov22}. See also Fig. \ref{figeps}. Thus, the Tauc-plot onsets 
more likely correspond to an indirect exciton but may also be influenced by defects. 
While the direct exciton gap is not expected to vary significantly with layer separation according to our
present calculations, because such excitons are a mixture of band to band transitions at different {\bf k}-points,
and because of the compensation of exciton binding energy and gap shift, 
the indirect gap exciton might have a somewhat higher binding energy and be more localized in {\bf k}-space.
For the bulk we obtain a lowest
direct gap at 4.2 eV and the lowest bright exciton is at 3.1 eV, indicating
an exciton binding energy of $E_B\approx1.1$ eV. Assuming that an indirect exciton associated with the indirect gap
of 3.8 eV has a similar binding energy, we would find
the optical indirect exciton gap in bulk at about 2.7 eV. This is still 0.3 eV 
larger than the onset of the Tauc plot in \cite{Reshma21}.
Thus we hypothesize that the exciton binding energy is larger for the indirect exciton.
Nonetheless, we might expect the indirect exciton to more closely follow a specific band edge and thus increase
slightly with increasing layer separation. Furthermore the nature of the indirect transition changes
to another {\bf k}-location of the VBM and the difference between direct and indirect quasiparticle
gap is reduced from that in bulk. Similar changes in direct/in-direct nature of the band gap going from the bulk to the monolayer limit are observed in several layered vdW systems~\cite{wu2019physical,acharya2021electronic}. We may thus expect that  for monolayers the optical gap even if still
indirect might approach more closely the direct gap exciton. 
Still our calculations of the indirect band gap shift between bulk an monolayer indicate
this shift would be of only  0.3 eV or so, which is significantly smaller than what is reported in \cite{Reshma21}.
To better understand this  discrepancy it will be necessary to calculate indirect exciton gaps and
to obtain a more detailed experimental analysis of monolayer optical properties.

\section{Conclusions}
In this paper we presented all-electron quasiparticle band structure calculations using a modified QS$G\hat{W}$ method, and optical response function calculations using the BSE approach.
The inclusion of ladder diagrams in calculating the polarization function
which determines the screened Coulomb interaction $W$ of the $GW$ method
is shown to reduce the self-energy correction of the gap beyond DFT by about a factor 0.77.  Our quasiparticle
band gaps
for the bulk  including these electron-hole effects are in good agreement
with the literature using a pseudopotential implementation but without
this electron-hole reduction of the $W$.  Because of this somewhat fortuitous
agreement on the quasiparticle gap, in spite of the different approximations
made in the calculation of $W$, we then find the excitons and imaginary part of the dielectric
function to also be in good agreement with prior work for bulk. There thus remains some
discrepancy on how to obtain the correct $W$ but once $W$ is established, good agreement is obtained in band structures and optical dielectric response.  Some effects of the strong local field effects in the direction
perpendicular to the layers were observed here and the appearance of
unphysical high energy sharp peaks was shown to be an artifact of the truncation   of the active space in the BSE. Finally,  the electronic screening only static dielectric constant was evaluated using an extrapolation from finite {\bf q} and found to give good agreement for the indices of refraction
with experiment to within about 15 \%. This confirms that in the QS$G\hat{W}$ approach both the band gaps and the screening are consistently in good agreement with experiment. 

For monolayers, we find an increased quasiparticle gap
but slow convergence of the quasiparticle gap
with the distance between the layers, as observed in other 2D systems.
On the other hand, the optical direct exciton gap converges much faster because as the
quasiparticle gap increases, so increases the exciton binding energy
because both are proportional to $W$, which is increased by reduced screening in a 2D system. The local field effect perpendicular to the layer were found to be even stronger in the monolayer than in the bulk.
While the direct gap at $\Gamma$ does not change much between bulk and monolayer at the GGA level, the top valence band becomes flattened out and this
increases both the smallest direct gap at $Z$ and the indirect gap.
Assuming a similar exciton binding energy for the  (not yet calculated) indirect
exciton as for the direct excitons,
we  predict a slight  increase of the optical absorption onset in monolayers.  
An increase in
optical gap was recently observed for exfoliated 2-3 layer thin samples
but was found to exhibit larger shifts than we here predict. 

{\bf Data Availability}: The data pertaining to various figures  are available at \url{https://github.com/Electronic-Structure-Group/v2o5-quasiparticle-exciton}. In particular,   XcrysDen\cite{xcrysden} (.xsf) and VESTA\cite{vesta}  (.vesta) datafiles related to Fig. \ref{figexpsi} $g-r$ are available  on \url{https://github.com/Electronic-Structure-Group/v2o5-quasiparticle-exciton} to facilitate  3D viewing.

\acknowledgements{This work was supported by the U. S. Department of Energy Basic Energy Sciences (DOE-BES) under Grant No. DE-SC0008933.  Calculations made use of the High Performance Computing Resource in the Core Facility for Advanced Research Computing at Case Western Reserve University. S.A. is supported by the Computational Chemical Sciences program within the U.S. DOE, Office of Science, BES, under award no. DE-AC36-08GO28308. S.A. used resources of the National Energy Research Scientific Computing Center, a DOE Office of Science user facility supported under award no. DE-AC02-05CH11231 using NERSC award BES-ERCAP0021783. We thank Vitaly Gorelov and co-authors of ref. 2 for providing the numerical data of their calculations and for useful discussions. We also thank Dimitar Pashov for his many contributions to the codes used here.}


\bibliography{Bib/lmto,Bib/dft,Bib/gw,Bib/v2o5,Bib/BSE,Bib/licoo2}

\appendix
\begin{widetext}
  \newpage
  \centerline{\bf Supplemental Information}
  \section{Discussion of sharp high energy features in $\varepsilon_2(\omega)$ in BSE}

\begin{figure}[h]
  \includegraphics[width=8cm]{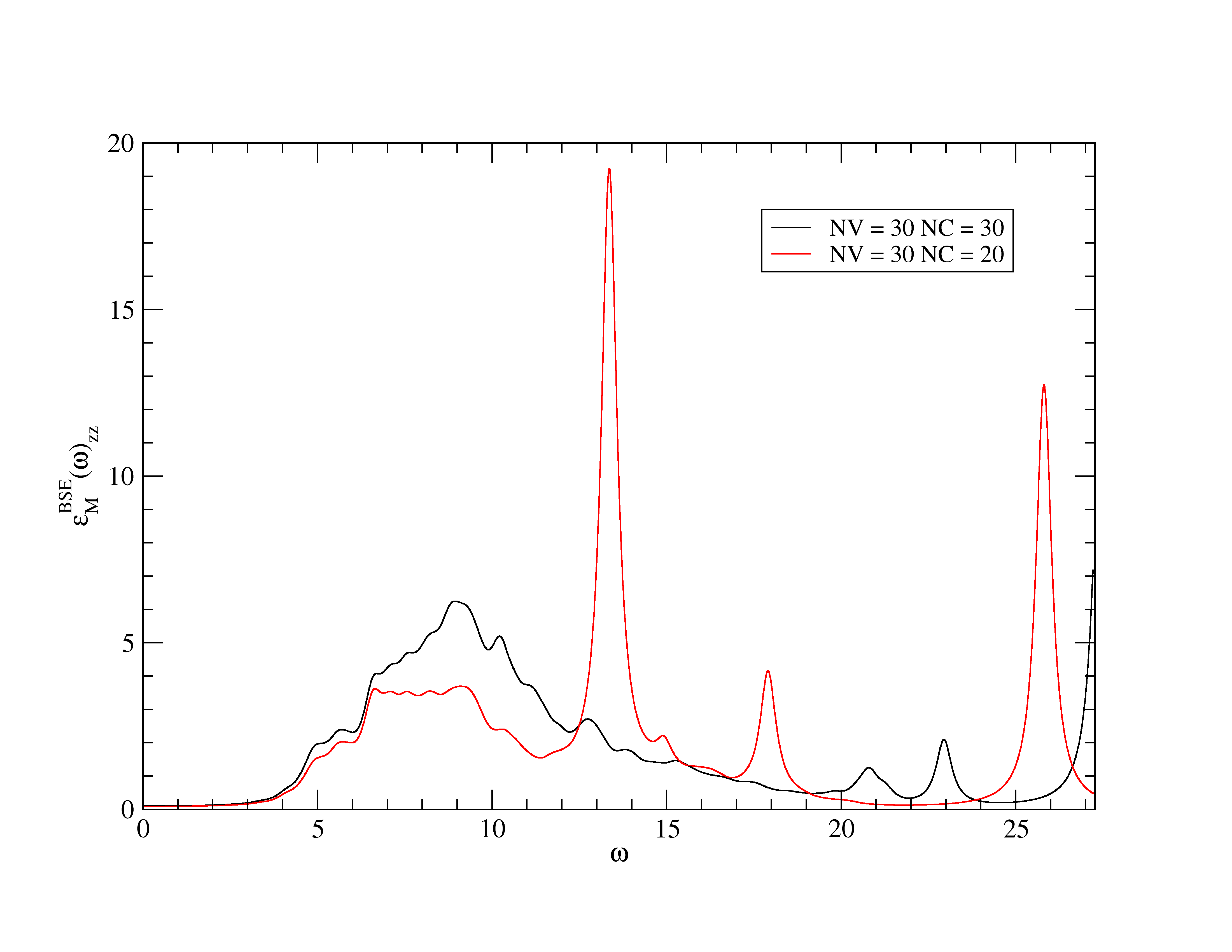}
  \caption{Macroscopic dielectric function for  polarization perpendicular to the layers obtained in BSE with different number of
    conduction bands.\label{fig:epsconverge} }
\end{figure}

\begin{figure}[h]
  \includegraphics[width=8cm]{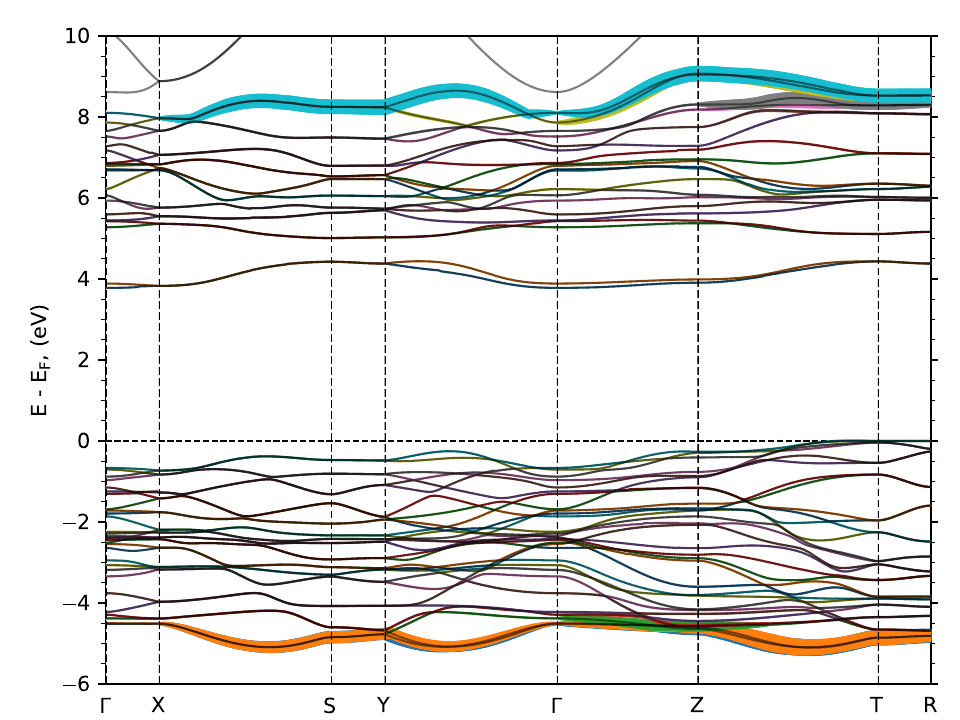}
  \caption{Bands contributing to the $\varepsilon_2(\omega)$ in a narrow energy range hear 1.5 eV where the sharp peak occurs in the BSE result.\label{figbandsharp}}
\end{figure}
\begin{figure}[h]
  (a)  \includegraphics[width=8cm]{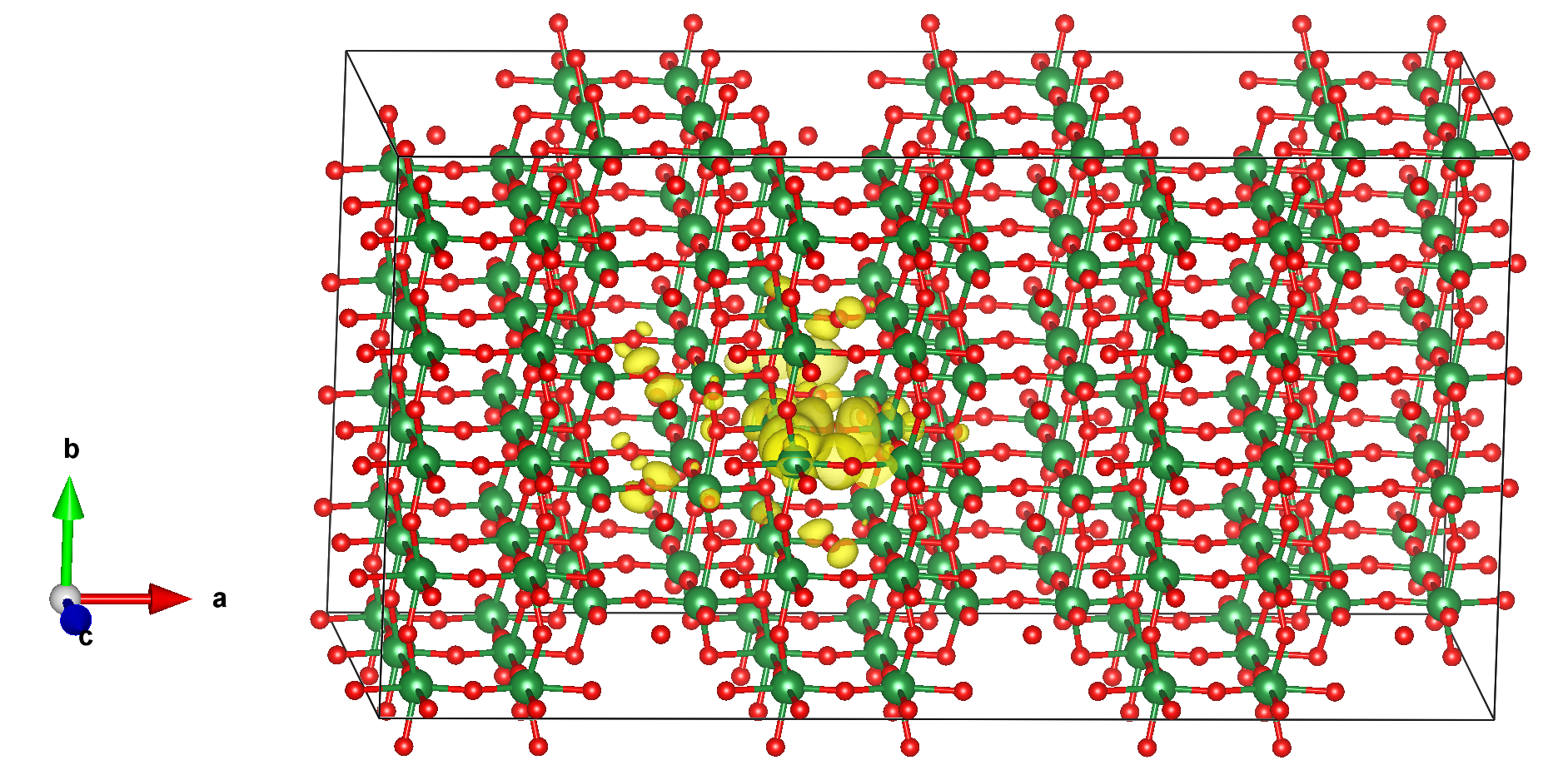}

(b) \includegraphics[width=8cm]{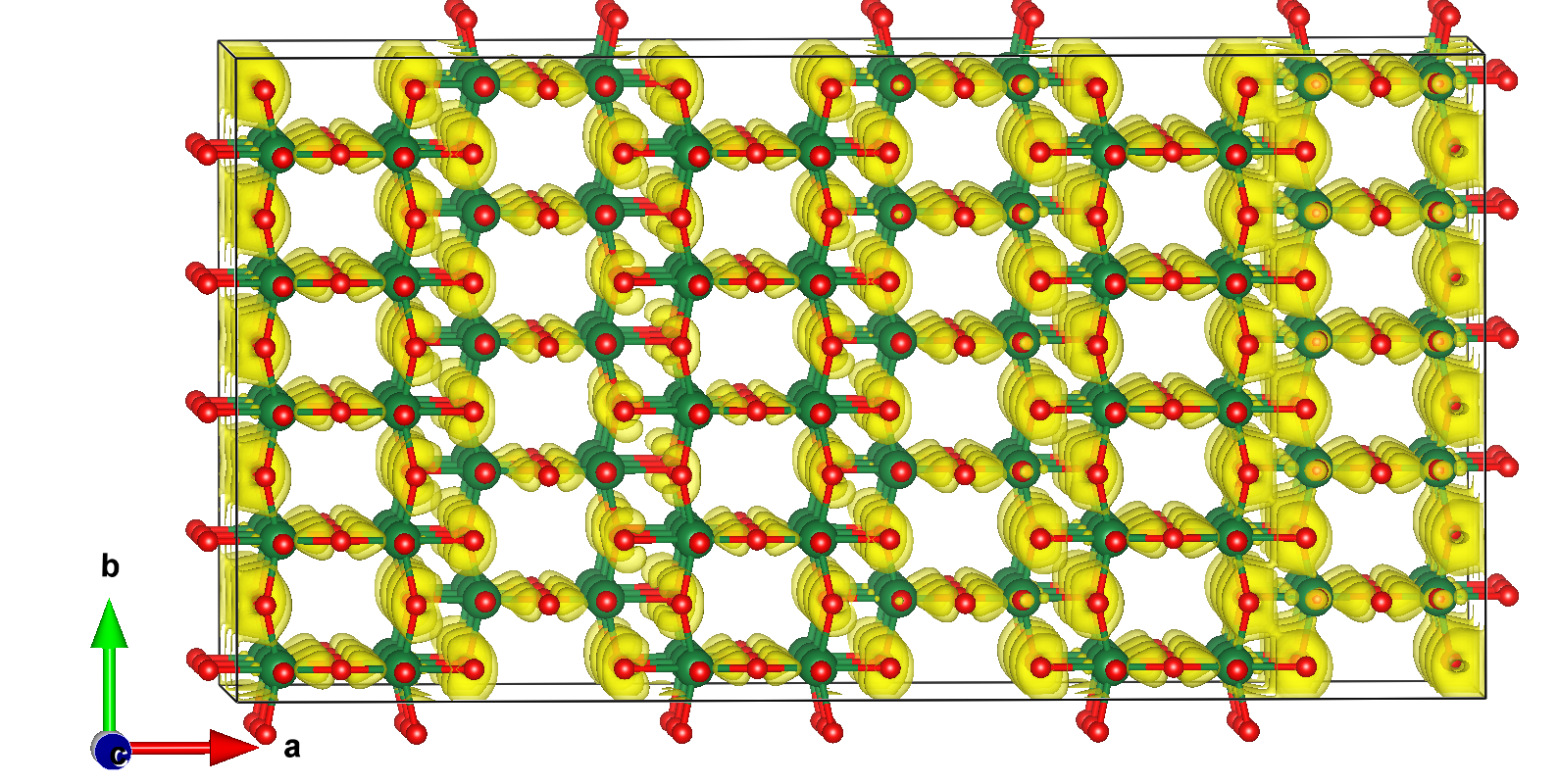}
  \caption{Two particle wave function for region near sharp peak at 13.5 eV showing the electron distribution
    when hole is placed on bridge oxygen in the center for (a) $N_c=20$
    and (b)  $N_c=30$.\label{figreal3020}}
\end{figure}

\begin{figure*}
  \includegraphics[width=14cm]{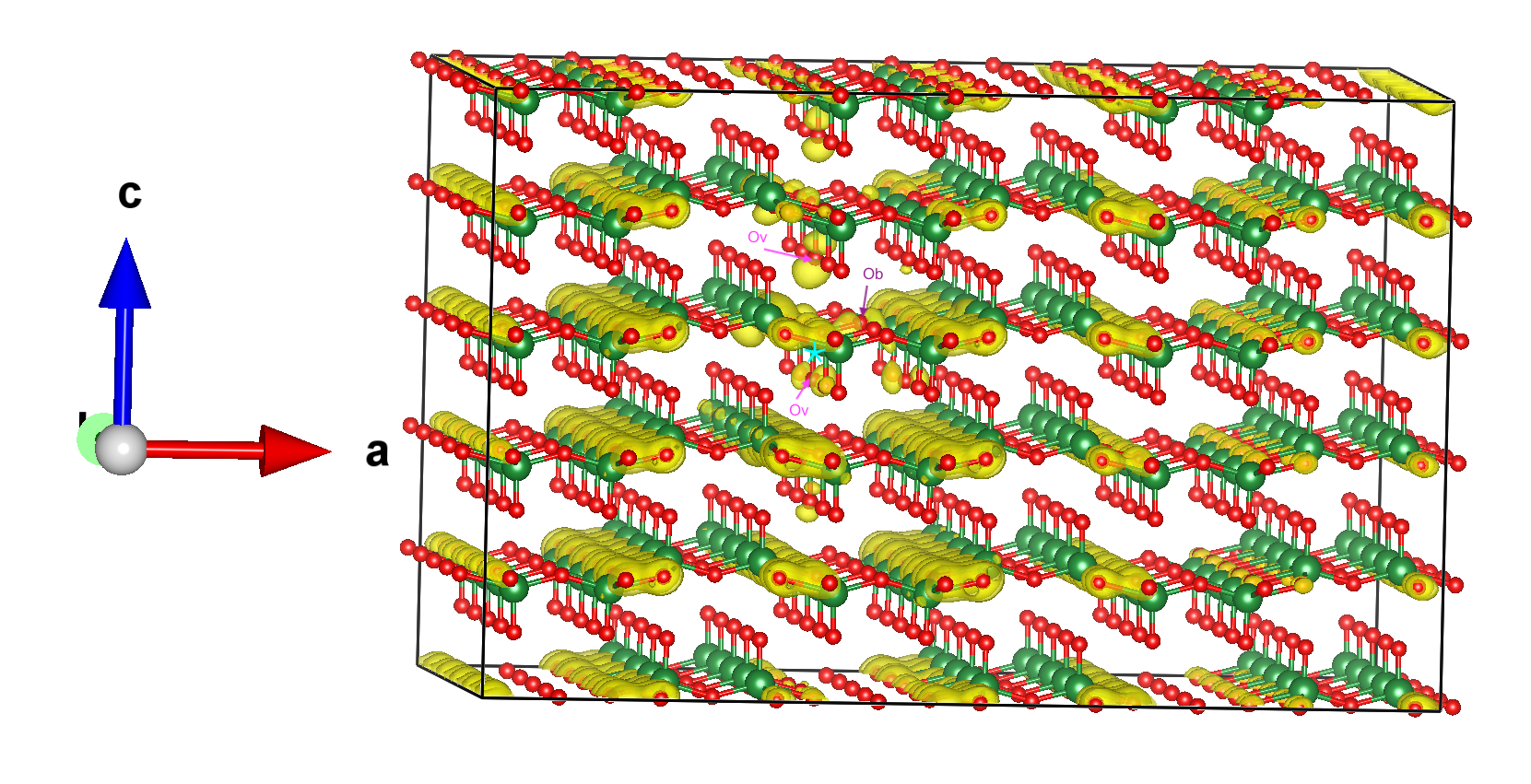}
  \caption{Two particle wave function for region near sharp peak at 13.5 eV showing the hole distribution when electron is placed on a vanadium
    for $N_c=20$ case. The V chosen is in the third layer from the top and in the center along the $b$ direction and is indicated
    by a light blue star. Some of its neighboring O atoms are labeled.
    \label{figv}}
\end{figure*}

In the main text in Fig. 3 sharp features appear at high energy for ${\bf E}\parallel {\bf c}$. That calculation was done with 30 valence bands and
20 conduction bands. Looking at an even wider energy range than in the main paper,  it appears that not only is there a feature at 13.5 eV but  peaks
also appear at 18 eV and 26 eV. 
If we include 30 instead of 20 conduction bands, the 13.5 eV and 18 eV  features disappear or at least are strongly reduced in intensity
as shown here in Fig. \ref{fig:epsconverge}.
However, we can see that the 26  eV feature is still there.   In the region below 10 eV the $\varepsilon_2(\omega)$ is somewhat
less suppressed indicating that some redistribution of oscillator strength from these high energy features to the lower energy region takes place by mixing more bands in the BSE active space.  Furthermore, when we examine which bands contribute to the
peak in $\varepsilon_2(\omega)$ in the energy range of the 13.5 eV peak, we find
it is strongly derived from the lowest O-$2p$ band and  the topmost V-$3d$ band. See Fig. \ref{figbandsharp}.
In  fact, within the active space of $N_c=20$, and $N_v=30$ these are the only bands which can give an energy band difference $\epsilon_{c{\bf k}}-\epsilon_{v{\bf k}}$ in this energy range. When 30 conduction bands are included, the band analysis still indicates
a strong contribution from the bottom O-$2p$ bands but  in the conduction band it is no longer  localized  near the top of the $d$-bands. Instead the weight is distributed over many bands. This spreading out of the oscillator strengths is related to the
transitions to strongly dispersing free-electron-like  bands above the
V-$3d$ conduction bands for $N_c>20$. This extreme dependence on the truncation of the active space of bands included, indicates that these sharp features at high energy are artifacts
of the truncation of the active space in the BSE calculation. 
To obtain reliable results  at increasingly higher energies would require one
to increase the size of the active space accordingly. For example, the peak near 26 eV also becomes reduced when
$N_v$ is increased  from 30 to 40 so as to include the O-$2s$ contributions. 

We also examine the changes between $N_c=20$ and $N_c=30$ for the two-particle distribution in real space.
When we place the hole on a bridge O$_b$ this  wave function is found to be fairly localized
to its V neighbors  and nearby O-V bonds as shown in
Fig. \ref{figreal3020}a for the $N_c=20$ case.  On the other hand, when $N_c$ is increased to 30, the wave function
becomes very delocalized as shown in Fig. \ref{figreal3020}(b). 

As mentioned in the main text, there is no compelling reason to focus on the hole at the O$_b$.
In fact, the bottom O-$2p$ bands near $-5$ eV  correspond to  $\sigma$-bonds from various oxygen types. 
Fig. \ref{figv}
shows that for the $N_c=20$ calculation, when we place the electron on V and examine
the hole distribution, not only the nearest bridge O$_b$ has a strong
contribution but there are also strongly localized contributions on the
vanadyl O$_v$, both the one bonded directly to the V on which the electron
is placed and the one in the adjacent layer. 
There are also quite delocalized contributions on the V-O$_c$ $\sigma$-bonds in
the double zigzag  chains. So, while this sharp peak  at 13.5 eV has some
localized aspects in terms of its two particle electron-hole wave function,
it also has delocalized aspects. In any case, it shows a strong inhomogeneity
in the direction perpendicular to the layers, which we can associate with
a strong local field effect. 

It remains curious that we only see these high energy features for the direction perpendicular to the layers, which suggests a  connection to  strong local field
effects. Inspecting the eigenvalues of the $H^{2p}$ two-particle Hamiltonians,
we do find eigenvalues up to $\sim$26 eV for the $N_c=20$ case, indicating
that the matrix elements  of the local field part of the kernel $\bar{V}$,
which are positive unlike those of $-W$ can significantly
increase the eigenvalues beyond $(\epsilon_{c{\bf k}}-\epsilon_{v{\bf k}})_{max}$ and lead to spurious peaks in the $\varepsilon_2(\omega)$. This indicates indeed  that the higher peaks at $\sim18$ and 25 eV 
are also related to local field effects. In the main text, we further substantiated this relation to local field effects based on the
similarity between $\varepsilon_2(\omega)$ and the loss function $-{\rm Im}[\varepsilon^{-1}(\omega)]$, which was pointed out in previous literature
to occur when strong local field effects occur. This indicates that the sharp peaks are in some sense plasmon-like corresponding
to a collective excitation of all the valence electrons included. However, when we truncate the BSE active space, these plasmons
have no chance to become damped by interaction with the electron-hole pair continuum in the energy above the band-to-band transitions
included. To obtain an accurate description of the dielectric function in this range, we need to expand the
BSE active space. The interaction between various electron-hole pairs is the key feature that  is included by the BSE.

\section{Extrapolating $\varepsilon_1({\bf q},\omega=0)$  to ${\bf q}\rightarrow0$}
Here we discuss some details of the finite ${\bf q}$ procedure to  calculate the static dielectric constant.
As mentioned in the main text, the bare Coulomb interaction  is replaced by a Thomas Fermi screened one to avoid numerical difficulties near ${\bf q}=0$.
The ${\bf q}\rightarrow0$ limit must be treated carefully because $W({\bf q})$
has an integrable divergence $4\pi/\varepsilon_M q^2$, which is handled
in the {\sc Questaal} codes by means of the offset-$\Gamma$ method and
by replacing the bare Coulomb interaction by $4\pi/(q^2+q_{TF}^2)$ with a small
$q_{TF}$. 
 Values of $q_{TF}^2$,  are chosen between 6$\times10^{-5}$ and $9\times10^{-5}$ and we then
 extrapolate  first $q$ to zero for different $q_{TF}$ and subsequently $q_{TF}$ to zero.  We find that for the $z$ direction
 the results depend more sensitively on the choice of $q_{TF}^2$ and hence a larger uncertainty results on the extrapolated results than for the $x$ and $y$ directions. Furthermore, using too small values of either $q$ or $q_{TF}$
 can  lead to unphysical results because of numerical artifacts.
 The extrapolation of the BSE results is shown in Fig. \ref{fig:epsq}.
 The top part shows the extrapolation as function of ${\bf q}$ for different
 $q_{TF}$ values and the bottom the extrapolation vs. $q_{TF}^2$.
 The data used in these plots and the linear extrapolation results are given in Table \ref{tabextrapol}. 
\begin{figure*}
 (a)  \includegraphics[width=13cm]{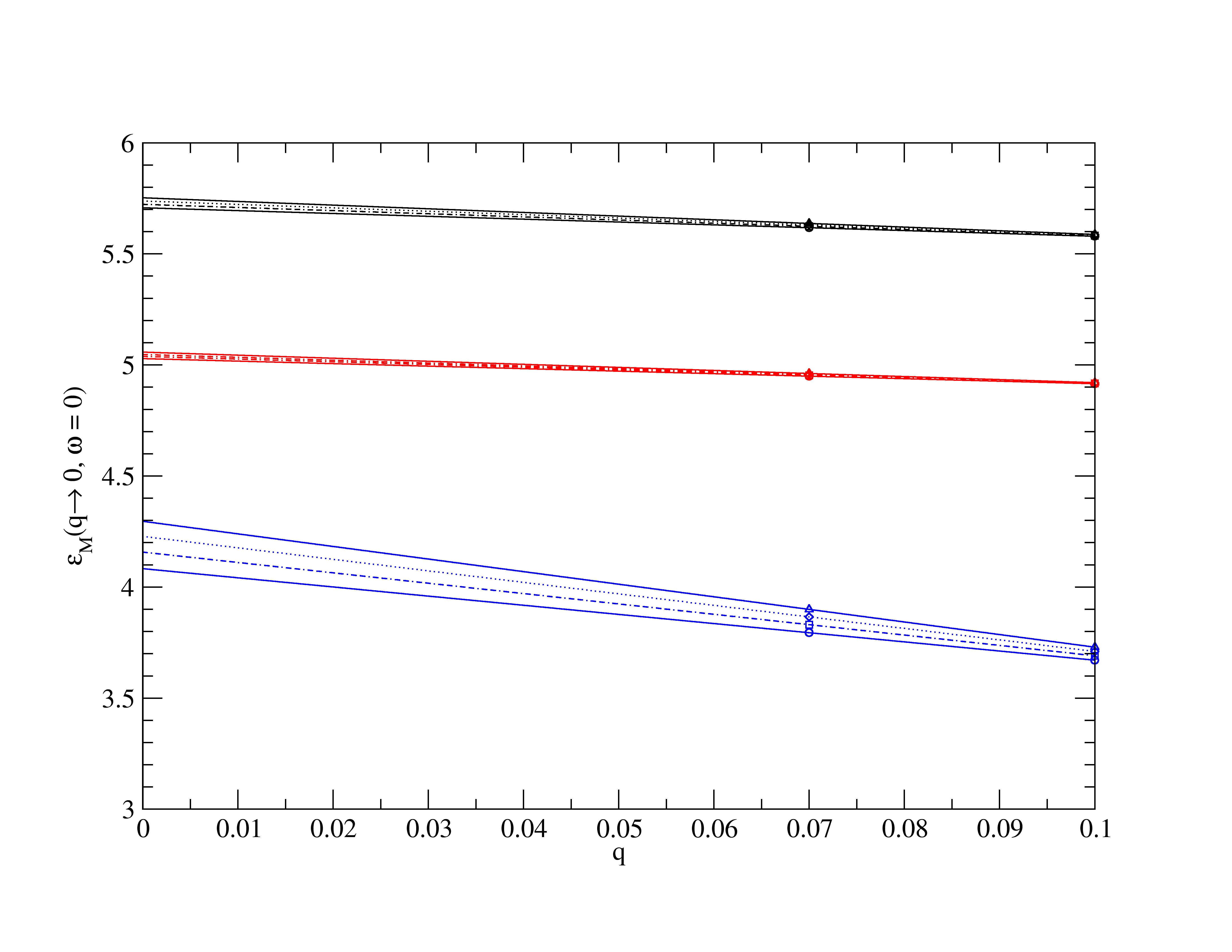}\\
 (b)  \includegraphics[width=13cm]{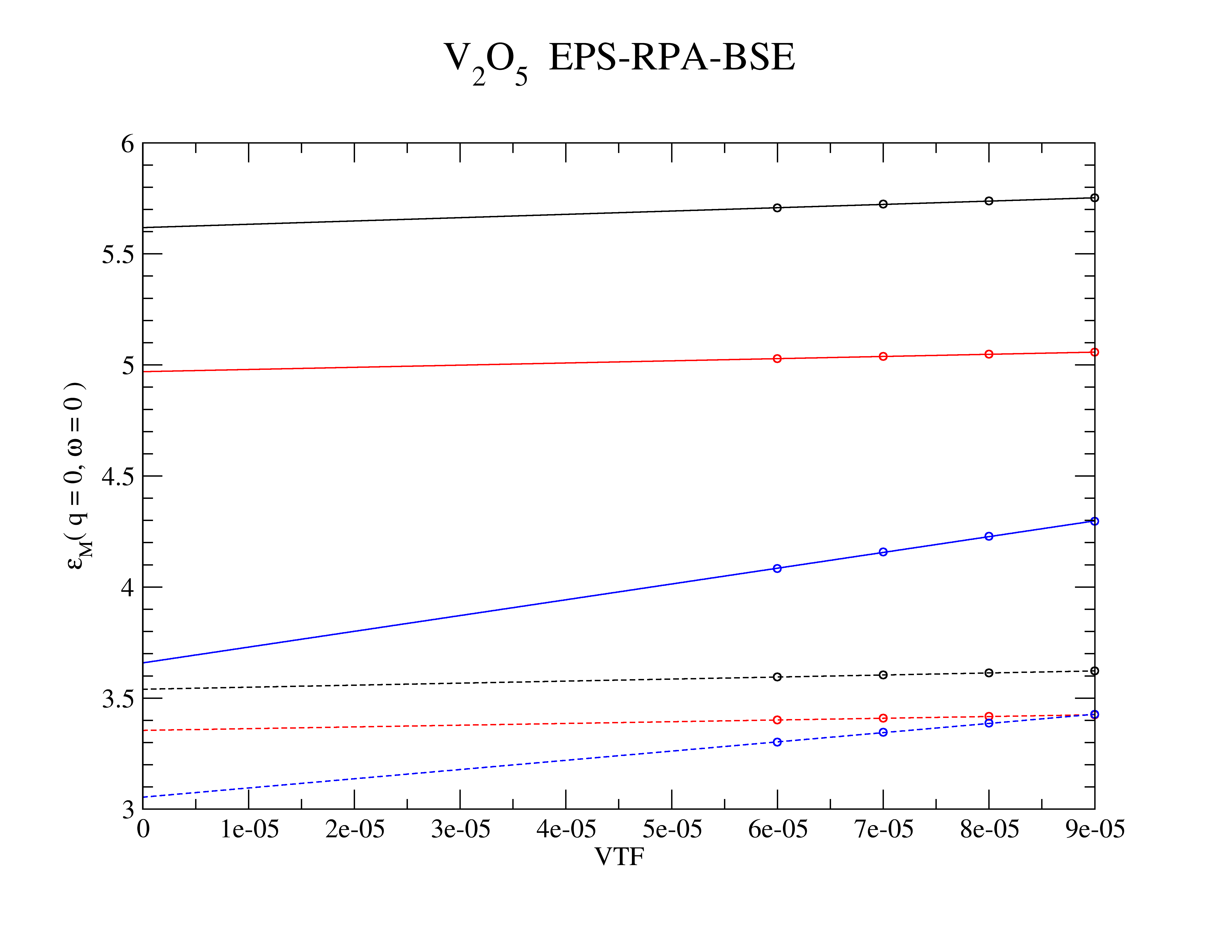}
    \caption{(a) Extrapolation of macroscopic dielectric function  $\varepsilon_M({\bf q},\omega=0)$ for ${\bf q}\rightarrow0$, obtained in BSE
    along different directions.
    Black curves correspond to ${\bf a}$ direction, red to ${\bf b}$ and blue to ${\bf c}$. The different symbols and the straight line interpolations through them correspond to different  values of the $q_{TF}$ parameter. The smaller  $q_{TF}$ corresponds to lower dielectric function at each $q$.
    (b) Extrapolation of $VTF=q_{TF}^2\rightarrow0$ of the ${\bf q}\rightarrow0$
    limits of $\varepsilon({\bf q},\omega=0)$ (solid lines) BSE, (dashed lines)
    RPA. 
    \label{fig:epsq}}
\end{figure*}
\begin{table*}
            \centering
          \caption{Dielectric constants $\varepsilon_1({\bf q}\rightarrow0,\omega=0)$ in RPA and BSE
            for various values of the parameter $q_{TF}^2=VTF$ and their extrapolation to $q_{TF}\rightarrow0$, given
            as $\varepsilon_\infty$. 
            In the last rows the index of refraction is obtained as $n=\sqrt{\varepsilon_1}$ and compared to experiment.} \label{tabextrapol}
	  \begin{tabular}{lcccccc}
		\hline
		\hline
	&\multicolumn{3}{c}{$\epsilon_M^{RPA}$($\bf{q} \rightarrow$ 0, $\bf{VTF}$, $\omega$ = 0)} &\multicolumn{3}{c}{$\epsilon_M^{BSE}$($\bf{q} \rightarrow$ 0, $\bf{VTF}$, $\omega$ = 0)} \\
		\hline
		\bf{VTF}& \bf{E}$||$\bf{a}  & \bf{E}$||$\bf{b} & \bf{E}$||$\bf{c}  & \bf{E}$||$\bf{a}  & \bf{E}$||$\bf{b} & \bf{E}$||$\bf{c} \\
		6.0e-5  & 3.5947&3.4014& 3.3019& 5.7071&5.0282&4.0831\\
	    7.0e-5  & 3.6044&3.4096&3.3457& 5.7235&5.0383&4.1577 \\
	    8.0e-5  & 3.6135&3.4174 & 3.3872& 5.7376&5.0480&4.2285\\
	    9.0e-5  & 3.6222&3.4248& 3.4264& 5.7521& 5.0575&4.2963\\
		\hline
		&\multicolumn{3}{c}{$\epsilon_M^{RPA}$($\bf{q}$ = 0, $\bf{VTF} \rightarrow$ 0, $\omega$ = 0)} 
		& \multicolumn{3}{c}{$\epsilon_M^{BSE}$($\bf{q}$ = 0, $\bf{VTF} \rightarrow$ 0, $\omega$ = 0)} \\
		\hline
		$\epsilon_\infty$  & 3.5400 & 3.3548 & 3.0541 &5.6183& 4.9698&3.6586 \\
		n$_{theo}$ & 1.88 & 1.83& 1.75 &2.37& 2.23&1.91 \\
		n$_{exp}$ & & &                          &2.07& 2.12&1.97 \\
		\hline
		\hline
	\end{tabular}

	\end{table*}
\section{Convergence of QS$GW$ self-energy.}
The convergnce of the QS$GW$ gap with {\bf k}-mesh is shown in Table \ref{tabgwk}. We choose the direct gap at $\Gamma$  to gauge the convergence of the
self-energy $\Sigma({\bf k})$ mesh. We can see that twith a $366$ mesh
the gap is converged to better than 10 meV and the {\bf k}-mesh
used in the main paper (155) differs from it by  only 47 meV.
\begin{table}
  \caption{Convergence of the direct gap at $\Gamma$ (in eV) with the QS$GW$ self-energy {\bf k}-mesh in bulk V$_2$O$_5$.  \label{tabgwk}}
  \begin{ruledtabular}
  \begin{tabular}{ccc}
    {\bf k}-mesh & $E_g$ & $\Delta E_g$\\ \hline
    133 & 5.098 &  \\
    155 & 5.074 & $-$0.026\\
    266 & 5.035 &  $-$0.039\\
    366 & 5.027  & $-$0.008  \\
  \end{tabular}
  \end{ruledtabular}
\end{table}
\end{widetext}
\end{document}